# Decisions at the end of the world

**Casper van Elteren**[*,1]

[*] *Corresponding author, casper.vanelteren@qut.edu.au*

[1] *Securing Antarctica's Environmental Future (SAEF), School of Mathematics, Queensland University of Technology, Brisbane, Australia*

**International treaties are often described as tigers without teeth, failing to secure compliance or change outcomes. Cooperation also depends on how states maintain the rules they have already made – especially where no sovereign stands above the states. Here we map more than six decades of decision-making under the Antarctic Treaty, linking submitted proposals to recorded responses and tracing how instruments act on earlier ones. Shared references tie decisions together across the agenda. Recurring revision concentrates in specialised histories, above all the management of protected areas, beyond what activity, popularity and reference age predict. Periodic administrative reviews retire outdated instruments in bulk, joining otherwise separate histories without coordinating policy across subjects. Binding Measures build overwhelmingly on earlier Measures, with little direct succession from non-binding Resolutions. Distinguishing these layers shows how a shared rule system is sustained by consensus, and provides a basis for comparing cooperation across global commons.**

## Introduction

Can a system of rules endure without a ruler to enforce compliance? There is one place on Earth that has escaped the traditional rise of nations. The region south of 60°S is often described as one of the largest global commons in the world and is governed through the Antarctic Treaty System. The Southern Ocean around it shapes global climate and sea level, so its governance concerns all of humanity (Rintoul, 2018). Its future is bound up with planetary issues such as climate change, biodiversity loss and resource scarcity (Steffen et al., 2015; Sutton et al., 2026). No state can solve these problems alone, nor keep them out by force (Biermann et al., 2012).

Signed amid Cold War rivalry and building on the scientific cooperation of the International Geophysical Year, the Antarctic Treaty set the continent aside for peace and science (Australian Antarctic Program, n.d.). The treaty froze the competing territorial claims without settling sovereignty. States keep their positions on sovereignty, but no new claims can be asserted while the treaty is in force (*The Antarctic Treaty*, 1959, Art. IV).

No single state rules Antarctica – yet for more than six decades, states have made and revised rules for it together, without resorting to violence. This cooperation is now being tested. With interstate conflicts in 2025 at their highest number since records began in 1946 (Uppsala University, 2026), China, with Russia's support, blocked the designation of the emperor penguin as a Specially Protected Species at the 2026 meeting in Hiroshima, although an overwhelming majority of parties supported the designation (Antarctic Treaty Consultative Meeting, 2026; Gibson, 2026). Under consensus, a single party can stall what nearly all others want. The system has nevertheless held, and its meetings have recorded proposals and decisions since 1961. Antarctica therefore lets us follow which rules states keep and which they amend, replace or retire.

Theories of international cooperation have long shown that repeated interaction, reciprocity and institutions can sustain cooperation between states without a central authority (Axelrod & Keohane, 1985; Keohane, 1984). Shared resources pose a harder problem. Hardin argued that pastures, fisheries and other commons are doomed to collapse when each user takes more than their share (Hardin, 1968). Ostrom showed that communities can avoid this fate by crafting their own rules and revising them as circumstances change, even without an authority above these communities (Ostrom, 1990). In these self-governing communities, compliance becomes quasi-voluntary. Users follow the rules because they trust that others will too, and monitoring one another keeps that trust credible (Ostrom, 1990). Ostrom's communities, however, were villages and irrigation districts in which households who lived off a resource could watch one another directly (Ostrom et al., 1999). Unlike these households, states are few but unequal, act on behalf of populations at home, and cannot expel a member that breaks the rules (Keohane & Ostrom, 1995). How such actors maintain a shared body of rules over decades is less well known. Many international arrangements keep long records, yet these are rarely read as histories of how one decision acts on another. A later decision may cite an earlier instrument without changing it, or may change or retire that instrument.

Whether cooperation between states succeeds is harder to say. A growing body of literature depicts many international treaties as tigers without teeth, failing to produce their intended effects (Hoffman et al., 2022). Ratifying human rights treaties has often not improved human rights practices (Hafner-Burton & Tsutsui, 2005; Hathaway, 2002). States that ratified the 1985 Helsinki Protocol on sulphur dioxide did cut their emissions, but the cuts cannot be attributed to the protocol itself (Ringquist & Kostadinova, 2005), and there is little evidence that the binding targets of the 1997 Kyoto Protocol reduced carbon dioxide emissions among the major industrialised emitters (Almer & Winkler, 2017). Antarctica has drawn similar criticism. Its decision-making appears increasingly unresponsive to rising environmental challenges, with binding agreements giving way to soft law (Gardiner et al., 2025), and the liability annex to its environmental protocol, adopted in 2005, has still not entered into force two decades later (Antarctic Treaty Secretariat, n.d.-a; Hemmings, 2018).

However, measuring the performance of a treaty is difficult. Effectiveness can mean solving the problem, changing behaviour or securing compliance, and each is hard to attribute to the treaty itself (Young, 2011). High compliance may simply reflect shallow commitments (Downs et al., 1996), and even counts of Antarctic agreements offer only a partial view of performance (Gardiner et al., 2025).

Teeth may also not be the only thing to measure. A standing forum gives states a place to exchange information, clarify obligations and resolve disputes through dialogue rather than force (Chayes & Chayes, 1993; Keohane, 1984), and communication alone can sustain cooperation (Ostrom et al., 1992). The forum also leaves states a body of rules to keep in working order. Such upkeep does not prove that a treaty works, since a rulebook can be tidy and still fail its aims, but it leaves a record that can be followed decision by decision. We therefore ask what states retain, revise and retire as successive decisions connect into a shared rule system, and where in that system they do most of their revising.

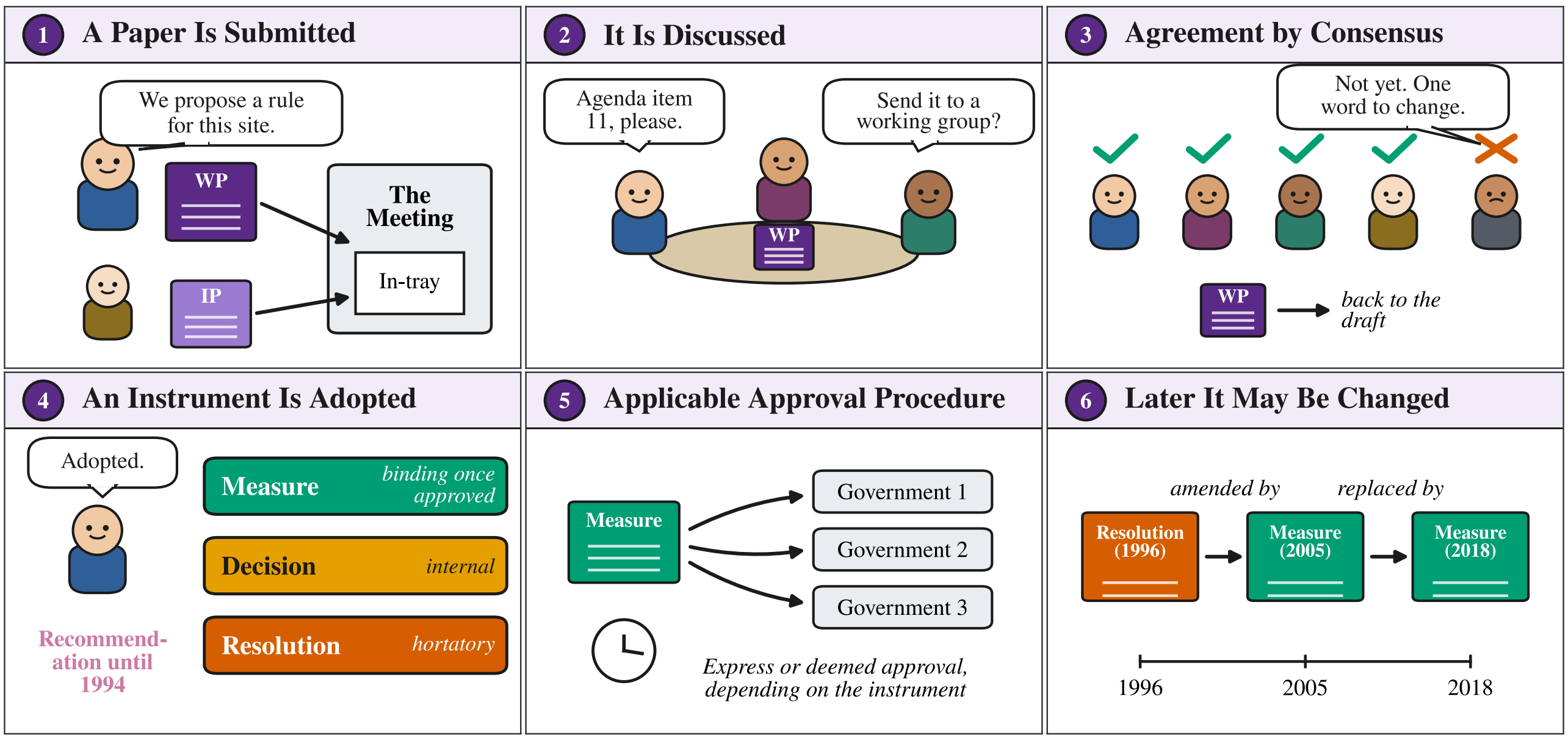

**Figure 1: From meeting procedure to linked decision histories.** A party or observer submits a paper for discussion, and the meeting needs consensus to adopt a numbered instrument. Since 1995, the forms are a Measure, binding once approved, a Decision on internal matters or a hortatory Resolution, while earlier meetings adopted Recommendations. Measures take effect under the applicable approval procedures, including tacit approval for management plans, and later meetings amend or replace the instrument. Agreement without a numbered instrument and other dispositions are distinguished in Figure S4.

Answering this requires knowing how a paper becomes an instrument and how later meetings act on that instrument. Parties and observers submit papers to the Antarctic Treaty Consultative Meetings (ATCMs), where since 1998 the Committee for Environmental Protection (CEP) has advised on environmental matters (Antarctic Treaty Secretariat, n.d.-b; c). The meeting discusses the papers and, when consensus is reached, may adopt an instrument. Since 1995, a Measure binds the parties once approved, a Decision concerns the meeting's internal matters and a Resolution is hortatory. Earlier meetings adopted Recommendations (Figure 1). Later meetings may amend, replace and revoke earlier instruments, so that a protected site remains an object of governance even as its management plan changes (Brunnée, 2002; Hogwood & Peters, 1982).

Earlier work has counted the inputs and outputs of Antarctic meetings and the range of topics they address (Gardiner et al., 2025), classified the policy acts of international organisations by topic and instrument (Lundgren et al., 2024) and shown how networks of legal texts reveal the evolution of rule systems (Coupette et al., 2021). None of these approaches separates a rule that cites an earlier one from a rule that changes the earlier instrument. Here, we map linked decision histories in the ATCM from 1961 to 2025, including the embedded CEP record from 1998. We link submitted papers to recorded responses and instruments, and follow how instruments act on their predecessors through amendment, replacement and revocation. These histories make the maintenance of shared rules measurable, so that later work can ask how it relates to the wider politics between states.

We find that decision-making under a common treaty combines specialised revision with periodic collective housekeeping. Protected-area plans and other recurring arrangements build long histories of succession, while most other connected histories remain short or linear. But administrative reviews connect these branches by declaring lists of older instruments spent, no longer current or withdrawn. References place 724 of the 793 instruments in one component, whereas actions on earlier rules place only 330, and 224 once the bulk retirement lists are set aside. References tie nearly every instrument into one framework, revisions link each plan to its predecessors, and housekeeping joins those branches when it retires old rules in bulk. A connected rulebook need not be a coordinated policy agenda. Antarctica lets us distinguish these forms of continuity in a system where every party can block a rule but none can impose one.

## Results

Decision-making is performed in the Antarctic Treaty Consultative Meetings (ATCMs), and its records are made publicly available through the Antarctic Treaty Secretariat after agreement of all Participants to the treaty. We link inventories of 6,368 papers and 793 numbered instruments through 1,429 distinct paper–instrument associations across ATCMs 1–47 (1961–2025) (Figure S6). The submitted documents contain concerns and issues that warrant discussion by the members, but need not end in a formal decision that affects governments. For example, in 2014 Australia proposed a checklist for assessing sites before clean-up, and the Committee for Environmental Protection agreed to add it to its Clean-up Manual without adopting a new instrument (Antarctic Treaty Consultative Meeting, 2014).

Of the 1,595 working papers submitted to ATCMs 16–47, 425 (26.6%) were endorsed or agreed without a numbered instrument at the meeting where they were submitted (Figure S4). The source-based classification records instrument adoption for 522 papers (32.7%). A further 286 (17.9%) were referred or deferred and 193 (12.1%) noted, while 41 (2.6%) record collective non-agreement. Because bargaining and objections often go unrecorded, this last figure cannot establish the frequency of disagreement or veto. Counting adopted instruments alone therefore captures only part of what the meetings agree on. These model-assisted classifications retain unclear outcomes and papers not identified in the report as separate categories (Methods).

## Connected by reference, fragmented by action

References bind almost the whole rulebook together. Recovered citations and recitals place 724 of the 793 instruments in a single component and only 25 stand alone (Figure 2 a). Amendments, replacements and revocations tell a different story (Figure 2 b). These 558 operative pairs join 549 instruments into 46 lineages, while 244 instruments have no operative link at all. The largest lineage, of 330 instruments, is dominated by protected areas (87% site-related), but bulk status reviews account for much of its size. Seven status-review Decisions retire older instruments in bulk, declaring them spent or no longer current, and contribute 129 links that cut across otherwise separate histories. Without them, the largest lineage shrinks to 224 instruments, the number of lineages rises to 48 and the number of isolates to 358. Most histories are small. Of the 46 lineages, 15 are pairs and 21 simple chains, while only ten branch or converge.

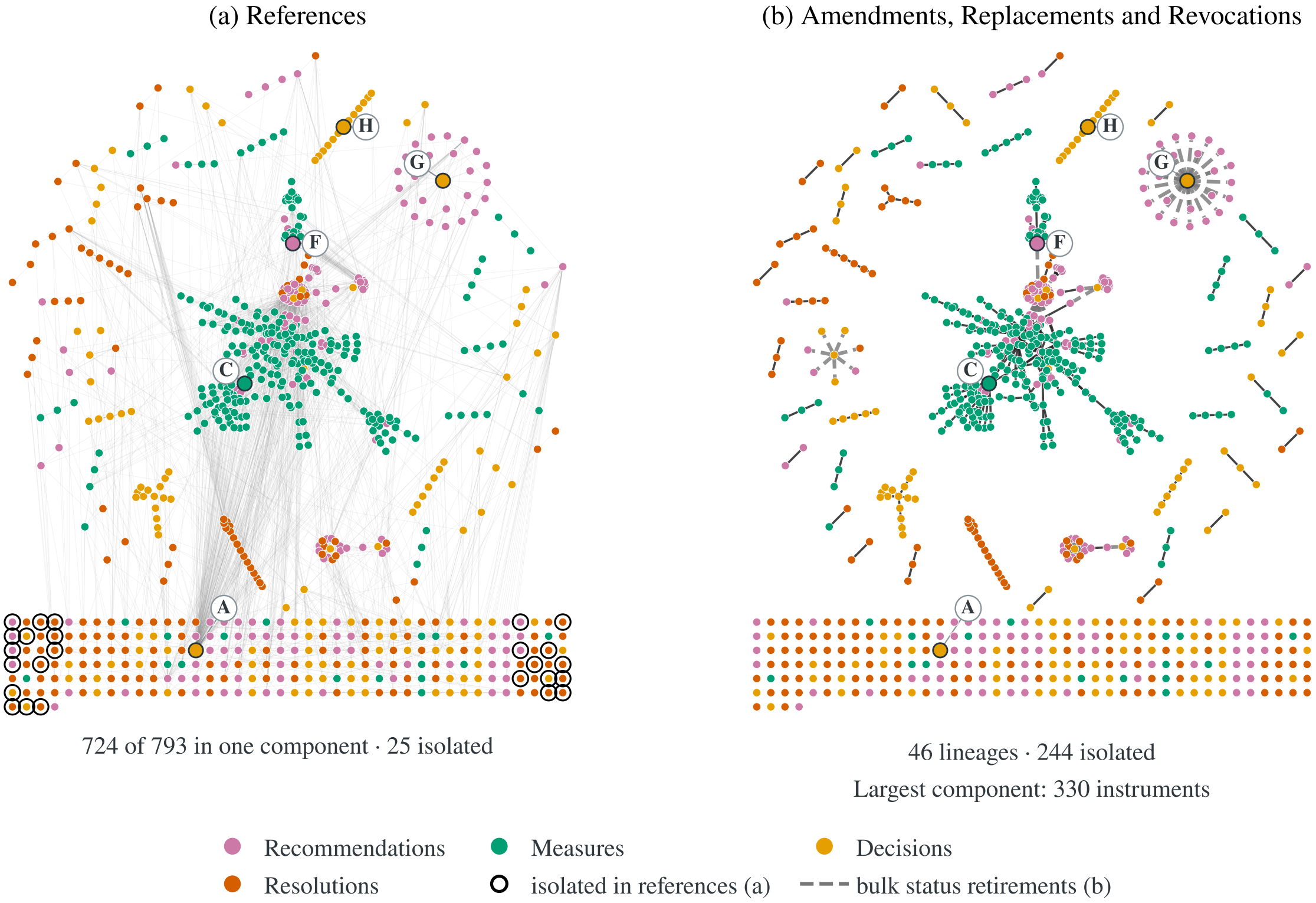

**Figure 2: Shared references, specialised revision and administrative connections.** The same 793 instruments occupy the same positions. Links run from the citing or acting instrument to its target, dated no later than the source. They are drawn without arrowheads, and components and lineages are weakly connected, counting links in either direction (direction is shown in Figure 4). (a) The reviewed documentary graph contains 2,782 distinct links. It places 724 instruments in the largest component, and 25 ringed instruments have no recovered link. Named-object projections are analysed separately. (b) The inclusive operative graph contains 558 pairs connecting 549 instruments into 46 lineages, alongside 244 isolates. Dashed grey links distinguish the 129 bulk status-review actions from the remaining operative links. Letters mark the same instrument in both panels. A is Decision 1 (2002), which renamed the protected areas and has 168 documentary links yet no operative action. C is Measure 2 (2005), a hub of protected-area plans. F is Recommendation VII-9, the root of the historic-site listings. G is Decision 3 (2002), a status review retiring 24 instruments at once. H is the 13-Decision strategic work plan. Lineage positions are computed from the operative graph. Operative isolates form the grid at the bottom, ordered so that each sits below its documentary neighbours. The grid is positional only in panel a. Removing bulk status reviews leaves 48 histories and a largest component of 224.

Removing bulk status reviews from both layers still leaves 700 instruments in the largest documentary component, against 224 for operative actions. A shared documentary framework thus coexists with rule change that proceeds in separate branches, periodically tidied by administrative review.

The distinction between citing and acting matters for reading the record. Of 943 recorded adoptions and designations of earlier instruments, 916 are preamble recitals recounting what those instruments did, not new acts (Figure S7 a to d). All 167 records concerning Decision 1 (2002), which renamed the protected areas, are such recitals. Treating them as actions would make plans appear to be revised again and again, and would join separate histories through widely cited system-level decisions.

## Revision follows built-in review

Where rules require states to return, histories accumulate. Annex V of the Environmental Protocol requires review of protected-area management plans to begin at least every five years, and the record follows this rhythm. Among site-plan Measures adopted in 1995–2012, 54% are first acted on after five years and a further 12% after six, and only 8 of 84 remain untouched within twelve years (Figure 3 a). Decisions and Resolutions show no such clock. Within the same window, 65.7% and 78.8% of those adopted in the period have no recovered action. Renewal routines elsewhere produce the longest series – 16 successive Resolutions on visitor site guidelines and 13 Decisions on the strategic work plan, each retiring its predecessor (Figure 4). Within these series, a new instrument typically replaces only its immediate predecessor while recalling several earlier versions. The recent visitor guidelines combine nine replacements with 34 recalls (Figure 4 c).

Most new instruments now act on earlier ones. The share of newly adopted instruments that amend, replace or retire a predecessor rose from 10% before 1995 (21 of 204) to 21% in 1995–2004 and 66% from 2005 (319 of 486, Figure S7 e). Protected areas account for much of this growth, but not all. Outside site rules, the share rose from 4% to 38% (84 of 224). Retirement lists also reach instruments that would otherwise appear forgotten. Of the 543 instruments adopted through 2014, 291 (54%) are replaced or retired at their first recorded action, and 228 (42%) have no recovered action by 2025 (Figure S7 f, Figure S5).

The binding form has narrowed in parallel. Of 279 Measures, 97% concern sites, and every Measure adopted since 2015 is a site rule. Since 2005, 226 of 251 Measures (90%) act on an earlier instrument, compared with 93 of 235 Decisions and Resolutions (40%), consistent with Gardiner and colleagues' account of the binding track (Gardiner et al., 2025). As annual output roughly doubled – from 10.8 instruments per year in 1995–1999 to 22.8 in 2020–2025 – a growing part of the meetings' work consisted of keeping existing rules up to date. Whether those updates broaden or narrow policy ambition is a question about content that the record of relationships cannot answer.

## Binding rules grow from binding rules

Binding and non-binding rules form largely separate tracks (Figure 3 b, c). Across all 558 operative pairs, 65.6% connect instruments of the same form. Leaving aside bulk retirements, the share rises to 84.1% (361 of 429), compared with 52.7% in networks matched for incoming and outgoing link counts, chronology and reference age (Figure S10). Measures, the only binding form, grow almost entirely out of earlier Measures. Of the 243 actions on a Measure, 239 come from another Measure. Of the 19 operative links between the forms used since 1995,

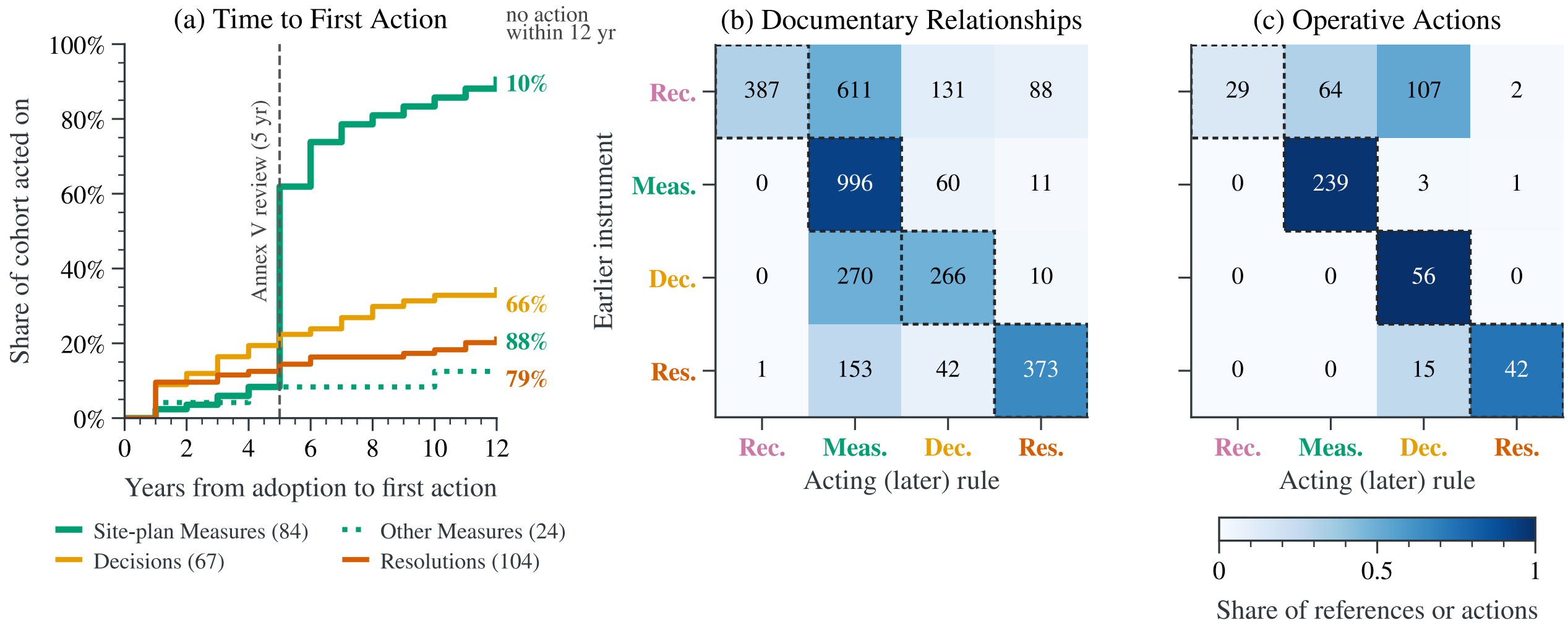

**Figure 3: Review timing and connections between instrument forms.** (a) Cumulative share receiving a first operative action within twelve years, for site-plan Measures (84), other Measures (24), Decisions (67) and Resolutions (104) adopted in 1995–2012. Percentages at right show no action within the window, not permanent inactivity. The dashed line marks five years, the Annex V review interval. (b) All 3,399 endpoint-eligible documentary records, before chronological filtering and including repeated observations. (c) The 558 unique operative pairs, including scoped replacement, retirement and approval requests. Rows identify the target instrument and columns the acting or citing instrument, and targets can be dated to the same year. Shading gives row shares, and dashed boxes mark within-form connections.

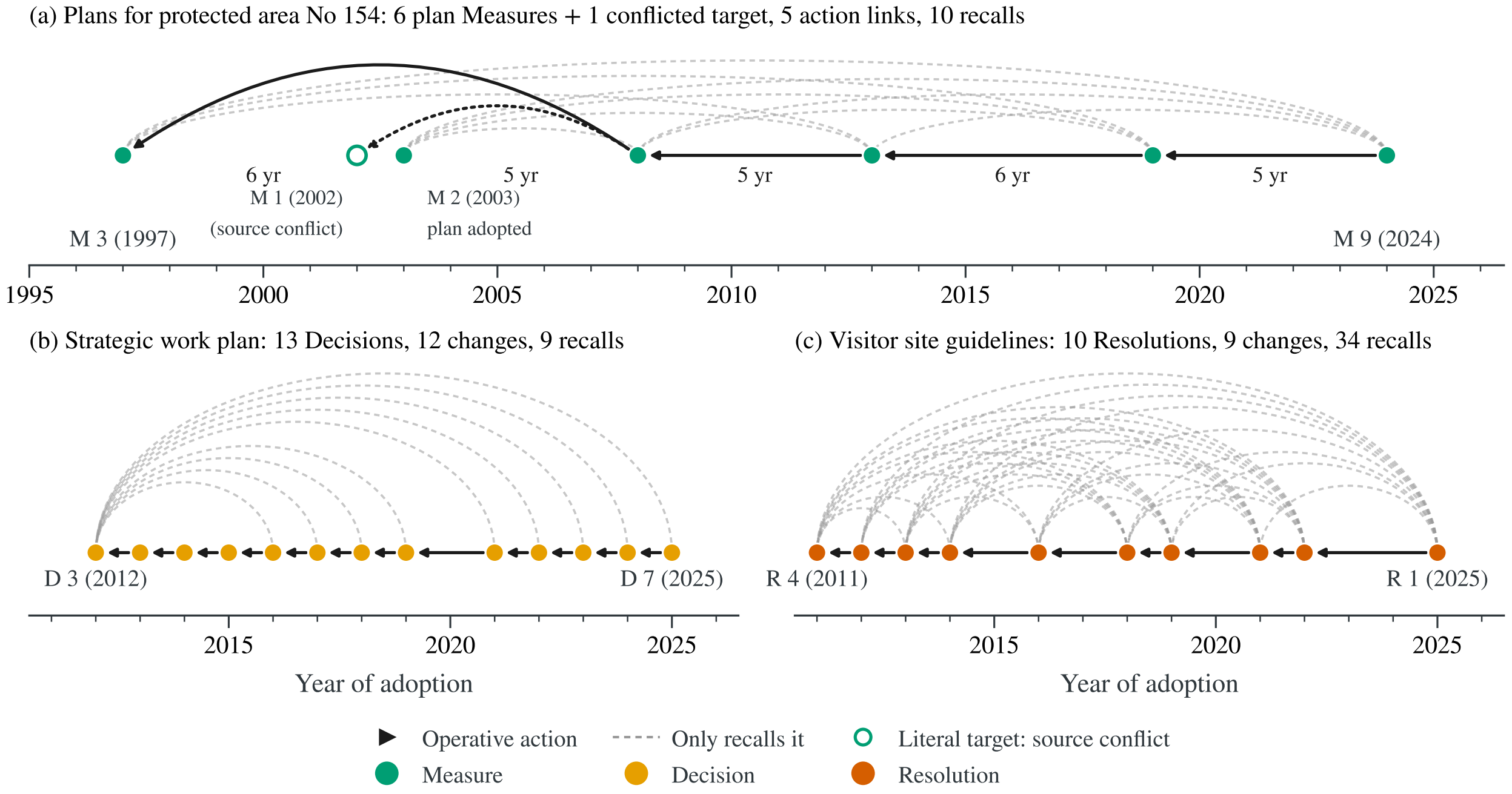

**Figure 4: Inside a silo, citation and rule change remain distinct.** Instruments are placed at their year of adoption. Solid arrows mark operative retirement or replacement, and grey dashed arcs mark references that only recall earlier instruments. (a) Six Measures adopting or revising management plans for Antarctic Specially Protected Area No 154, plus a target named in a conflicting source citation. Measure 2 (2003) adopted the plan and is recalled by four later Measures. The 2008 operative clause instead names Measure 1 (2002), whose original adoption list does not include ASPA 154. That literal target is retained as a hollow node with a dotted arrow and is not silently reassigned to 2003. Five operative links, including this conflicted citation, and ten recalls are displayed. Intervals refer to the six plan instruments. (b) The strategic work plan, with 13 Decisions and 12 successive retirements. Nine direct recalls of the 2012 Decision are displayed as dashed arcs, and the first action, in 2013, also cites 2012. (c) Visitor site guidelines since 2011, with ten Resolutions linked by nine retirements and 34 recalls, the latest part of the longest recovered chain (16 Resolutions). Two references in the 2016 recital are resolved to Resolutions from context despite the source's incorrect "Recommendations" label. Model-assisted source review does not constitute independent human validation.

17 are bulk retirements, one updates administrative fields and one requests approval of a Measure. None turns a hortatory Resolution into a binding Measure. Older Recommendations, the single form used before 1995, supply most of the remaining links between forms (173 of 192). References show somewhat more contact. Measures refer to or act on 23 distinct Resolutions, and 11 of 174 Resolutions refer to or act on a Measure. Most of these records (90 of 153), however, recall two Resolutions from the 1990s that extended the expiry dates of protected sites.

The separation also tracks what instruments do. Within the six title-derived function classes, 93.5% of non-bulk operative pairs remain in the same class, compared with 73.3% when the null also preserves each actor's target-form pattern. Conversely, holding function patterns fixed raises the expected share of within-form connections to 80.9%, close to the observed 84.1%. Functional specialisation therefore persists beyond form mixing, while much of the form separation overlaps with instrument function. These classes describe recurring instrument functions rather than independently coded policy agendas.

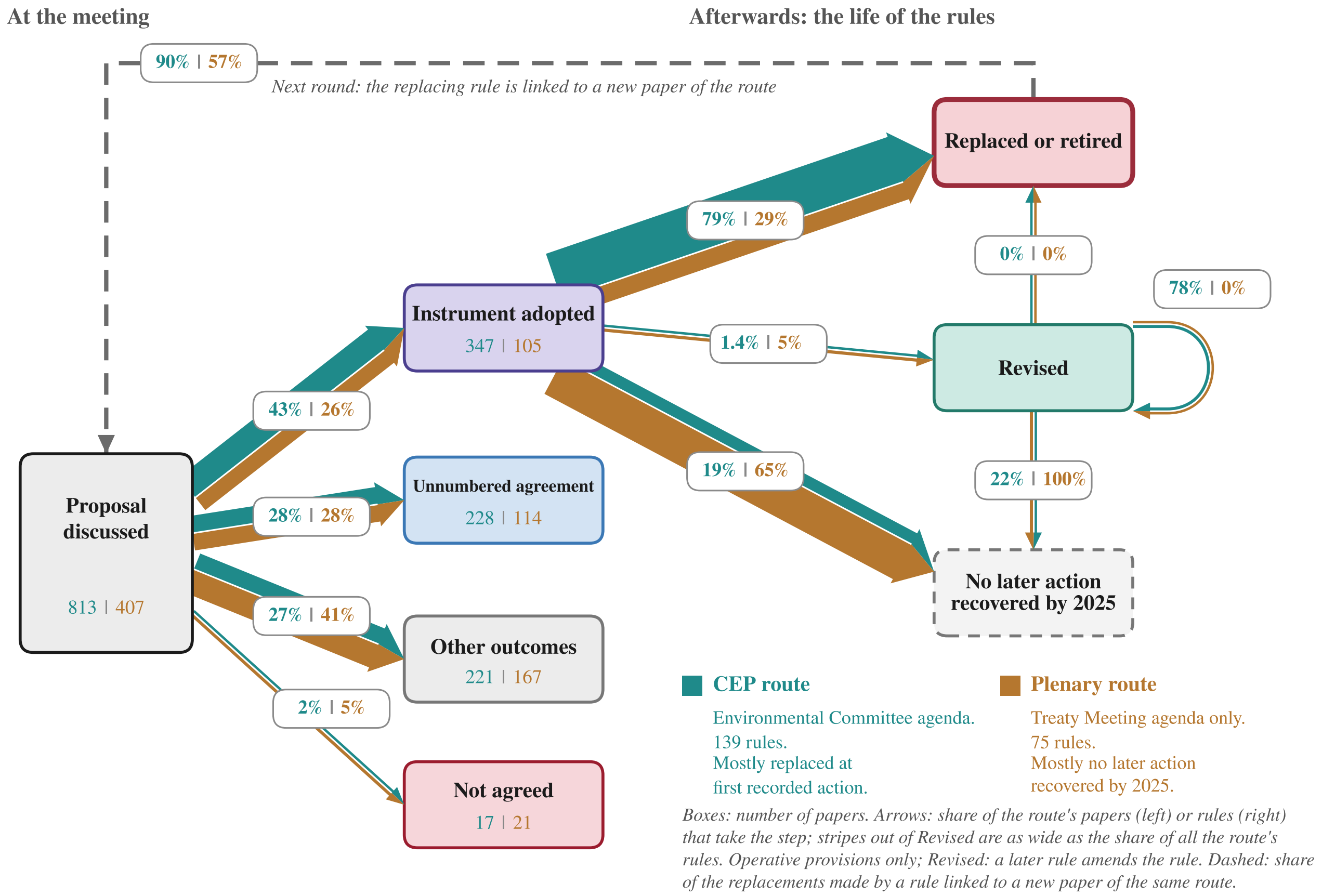

**Figure 5: Agenda routes and subsequent instrument histories.** The left side shows report-based dispositions for discussed working papers of ATCMs 22–47 assigned only to the CEP agenda (teal, 813) or only to the plenary agenda (brown, 407), omitting 138 papers without an exclusive route. Instrument adoption requires a recorded connection between the paper and collective adoption of a numbered text. The Other outcomes category groups noting, referral, deferral, withdrawal and unclear dispositions. Not agreed denotes collective non-agreement and does not independently establish a veto. The right side shows subsequent operative actions on associated instruments adopted in 1998–2014 (139 CEP, 75 plenary), followed through 2025. The two sides use different units and cohorts, so the arrows do not give a single probability of a proposal becoming an instrument. Revised denotes amendment by a later rule, and replacement includes component replacement and administrative retirement. Arrow shares use papers on the left and rules for the first step after adoption. Stripes leaving Revised are scaled to all the route's rules. No later action means none recovered by the endpoint. Dashed return links count replacement-or-retirement pairs whose acting instrument is associated with a new paper of the same route (143 of 159 CEP pairs, 90%, and 13 of 23 plenary pairs, 57%). Stronger and shared-subject associations enter route assignment and return links. These associations do not establish proposal adoption or causal contribution. Route, instrument form and subject overlap, but alternative route attributions retain the replacement contrast.

## Committee preparation accompanies turnover

The committee structure of the meeting maps onto these histories (Figure 5). Among working papers from ATCMs 22–47 assigned to a single agenda, 813 went to the Committee for Environmental Protection and 407 to the plenary. Instruments associated with CEP papers are far more often replaced or retired (79% of the 139 adopted in 1998–2014) than those associated with plenary papers (29% of 75). The two portfolios differ in kind. Almost all route-associated Measures (112 of 116) come through the CEP, and most Decisions (33 of 37) through the plenary. Even outside site rules, CEP-associated instruments turn over more often (13 of 22, 59%) than plenary ones (22 of 75, 29%). Replacements also feed back into the agenda. In 90% of CEP replacement pairs (143 of 159), the acting instrument is associated with a new CEP paper, against 57% for the plenary (13 of 23). The contrast holds under stricter link definitions and a later cohort (76–85% against 26–29%), but route, form and subject overlap too closely to isolate the effect of committee preparation itself.

## Discussion

We expected six decades of decisions under a single treaty to form a connected body of law. References place 724 of the 793 instruments in one component, but this shared framework contains different kinds of connection. Recurring plans and guidelines build specialised histories of revision. Periodic status reviews reach across those histories, declaring whole lists of older instruments spent, no longer current or withdrawn. They enlarge the largest operative component from 224 to 330 instruments. Citing a rule, changing it and retiring it are different acts. A map that merges them makes the system look more integrated than its routines of revision are. Shared references, specialised revision and collective housekeeping hold the Antarctic rulebook together (Figure 2).

The distinction also changes what an isolate means. Five recent Measures designate new protected areas that have no earlier plan to replace. They stand alone in the operative graph, yet their annexes cite earlier guidance and instruments. Source review recovered such references and reduced documentary isolates from 42 to 25. Standing outside a lineage is therefore not the same as standing outside the rule system. A new object of governance and a missed citation can both look like isolation.

## Where lineages come from

Some decisions form long histories while others stand alone, depending less on what a decision is about than on how its rules are designed. The Treaty's articles define its scope, jurisdiction and procedures (*The Antarctic Treaty*, 1959). They anchor the framework and remain outside chains of succession. Lineages form where rules or routines require states to return. Annex V requires review of protected-area management plans to begin at least every five years (Antarctic Treaty Consultative Meeting, 1991, Art. 6(3)), and first actions cluster after five and six years (Figure 3 a). The strategic work plan accumulates 13 successive Decisions, and visitor site guidelines a still longer chain of Resolutions. The division of labour within the meeting points the same way. Instruments associated with CEP papers are replaced or retired more often than those from the plenary, 79% against 29% by 2025 (Figure 5). Route, form and subject overlap, so the record cannot isolate the effect of the committee itself.

## Maintenance, teeth and safe ground

Most of this revision concerns the management of places. Even with bulk status reviews set aside, 334 of the remaining 429 operative links connect site-related instruments. Binding activity may have narrowed into routine upkeep – the hallmark of a tiger without teeth (Gardiner et al., 2025). Or clear objects and scheduled returns may be exactly what allows cooperation to endure. The graph shows where maintenance concentrates, but it cannot measure the stakes or ambition of each revision.

The emperor penguin case shows what a record of adopted decisions leaves out. The proposed listing as a Specially Protected Species was blocked despite broad support (Antarctic Treaty Consultative Meeting, 2026; Gibson, 2026). Under consensus, disagreement can halt one decision while other work continues. Reading lineages alongside stalled proposals and agenda access would show whether revision avoids contested ground – a question this record raises but cannot answer.

## Hard and soft law in parallel

Legal forms also divide the rule system. Measures build almost entirely on Measures. Of the 243 operative actions on a Measure, 239 come from another Measure. Of the 19 direct links between forms adopted since 1995, 17 are bulk retirements, one updates administrative information and one requests approval. None converts a non-binding arrangement into a binding one. Soft law is often understood as a way to test arrangements before hardening them into commitments (Abbott & Snidal, 2000). In Antarctica, we find little sign of that path. References show more contact – Measures cite or act on 23 distinct Resolutions, and 11 Resolutions cite or act on a Measure – so influence between forms may run through less explicit channels. The conditional comparisons show that much of this separation overlaps with instrument function, but they do not isolate an effect of binding status.

## Revising rules and retiring the archive

Restoring the retirement lists overturns the record's first impression that most older decisions are never revisited. Among instruments adopted through 2014, only 228 of 543 have no recovered action, while 291 reach replacement or retirement at their first recorded event. Retirement and revision are distinct acts. An instrument can become obsolete long before a meeting records it as spent, and a replaced plan can leave other provisions of its instrument intact. New arrangements accumulate alongside older ones – what institutional theory calls layering (Mahoney & Thelen, 2010; Streeck & Thelen, 2005) – but neither an isolate nor a late retirement date proves layering. The pattern is rather a division of institutional work. A healthy democracy must reconcile two competing demands – the stability of its foundational principles and the capacity to adapt to the changing needs and values of its people (Elkins et al., 2009). Its courts and legislatures can make the outcome stick. The Antarctic Treaty System has only the consensus of the states that established the system. Its answer to the same demands is to keep its framework fixed, revise a few branches continually and clear the inherited archive from time to time.

## Where the machinery will be tested

This division of labour may suit the continent's current position. Much of Antarctica's value remains latent. Mineral resource activities other than scientific research are prohibited (*Protocol on Environmental Protection to the Antarctic Treaty*, 1991), territorial claims are frozen, and the climatic role of its ice and ocean belongs to no state in particular (Rintoul, 2018). Its future tests lie beyond the upkeep of sites. A binding framework for tourism remains under discussion, and the protection of a climate-threatened species has stalled (Ministry of Foreign Affairs of Japan, 2026). These domains show far fewer revision histories than protected places and visitor guidelines (Figure S9), although administrative retirement can draw a domain into a large component without any routine of substantive revision. The comparison identifies areas for further study without predicting where governance will fail. Joining decision histories to proposals, rejected drafts and deferred negotiations would distinguish a maintained branch, an obsolete rule and an unresolved demand for action.

## Towards comparison

We also ask what Antarctica can teach about governing without a sovereign more broadly. Ostrom observed that durable commons organise their activities in multiple, nested units, each handling its own problems within a shared framework (Ostrom, 1990). The Antarctic record shows a related structure among states, with a common framework to which most decisions refer and separate branches in which rules are maintained. Such compartmentalisation may help explain how cooperation survives disagreement. At the same Hiroshima meeting that failed to protect the emperor penguin, the parties established two new Antarctic Specially Protected Areas (Ministry of Foreign Affairs of Japan, 2026). The coexistence of these outcomes shows that disagreement on one issue did not stop all adoption. The forum also remains a channel for dialogue where agreement fails (Chayes & Chayes, 1993). The network alone cannot explain why those outcomes coexist, and one case cannot establish that this structure causes resilience. It does, however, offer a way to examine a question with which treaty research has long struggled. As well as asking whether a treaty has teeth – whether it secures compliance or measurable outcomes (Hoffman et al., 2022; Young, 2011) – one can ask how its decisions build on one another, which branches are maintained, which stand still and where the record connects. Other regimes that govern shared resources by consensus, such as the Commission for the Conservation of Antarctic Marine Living Resources (Brooks et al., 2020), or that accumulate decisions through conferences of the parties (Brunnée, 2002), leave comparable records. Comparing how they revise their rules would show whether the Antarctic pattern belongs to a continent of latent value or is a general feature of consensus governance.

## Limitations

Our map shows only what entered the record. The reports document proposals that reached the table and the decisions taken on those proposals. Proposals that were never tabled, informal bargaining between meetings and the setting of the agenda itself leave no trace, yet such non-decisions can shape governance as much as recorded ones (Bachrach & Baratz, 1963). The emperor penguin case is visible precisely because the proposal was tabled and blocked. Much of what does not happen in Antarctic governance remains invisible to any documentary approach. Collective non-agreement, recorded for 2.6% of working papers, therefore cannot be read as the rate at which consensus blocks action.

The reconstruction also rests on automated reading of the reports. Extraction errors can alter individual histories and, through them, aggregate comparisons. Historical wording may be mistaken for a current action, and genuine actions may be missed (Methods). Coverage changes over time. Early reports often name topics and delegations without identifying papers by number, so the analyses of papers and agenda routes use later cohorts, while instrument histories extend to 1961. Lineages depend on aggregation, and a single component need not represent one continuous policy. The design describes one forum and cannot identify the effects of consensus, committee preparation or approval procedures. Bargaining outside the reports, national implementation and environmental outcomes remain beyond the record. These limits prevent us from inferring governance effectiveness from recorded activity, while leaving the structure of rule revision open to examination.

**Table 1: Relations retained in a linked decision history.** Identifiers, dates, evidence passages and extraction provenance remain attached to the records. An unnumbered agreement can have a disposition without an instrument node.

| Record | Observation | Interpretation limit |
|---|---|---|
| Paper outcome | Source-assessed disposition at the submission meeting. | Recommendation-level outcomes can differ, and an instrument association alone does not establish adoption. |
| Paper–instrument association | Named proposal, contribution or shared subject, with timing and evidence. | A shared subject does not establish proposal adoption or causal influence. |
| Operative relation | An operative provision addresses an instrument dated no later than the acting instrument. | Replacement can change policy. A request for approval is not approval. |
| Recital or citation | An instrument recalls an earlier act or cites a founding document. | Recall does not itself change the earlier instrument or establish its legal status. |
| Named-object reference | An instrument names a code, resource, declaration or institutional arrangement. | A connection through an object is distinct from a direct citation to its associated instrument. |

## Conclusion

Can a system of rules endure without a ruler to enforce compliance? For more than six decades, the states that govern Antarctica have kept one alive. Mapping their decisions reveals the shape of that system, which combines a shared documentary framework, specialised histories of recurring revision and periodic reviews that retire older rules across those histories. This structure accumulated one consensus decision at a time, without an overarching design. These connections do different work, and keeping them apart prevents a tidy rulebook from being mistaken for a coordinated agenda or effective environmental protection. As planetary problems press on institutions built for sovereign states, how states maintain their shared rules deserves as much attention as whether treaties have teeth. Antarctica offers a place to observe rules without rulers – and a framework for comparing how other regimes keep their decisions alive.

## Methods

### Corpus and linked decision histories

We analysed the English-language final reports of ATCMs 1–47, including the embedded CEP reports, alongside the Antarctic Treaty Secretariat's instrument and submitted-paper inventories (Antarctic Treaty Secretariat, n.d.-d). The instrument inventory contains 793 numbered Recommendations, Measures, Decisions and Resolutions, including numbered special-meeting instruments whose available texts were read separately. The paper inventory contains 6,368 working and information papers consolidated by meeting, type and number. Revisions and language editions do not become additional papers. We identify instruments by type, number and adoption year, or by the historical meeting–sequence designation for Recommendations, so that Measure 1 (1995) and Measure 1 (1997) remain distinct instruments.

A linked decision history contains several kinds of evidence. Paper outcome records retain source-assessed dispositions independently of instrument associations. Paper–instrument associations record a proposed instrument, a documented contribution or a shared subject, with the assigned label and its evidence. Operative relations record amendment, replacement or revocation, or requests that Parties approve an identified instrument. Founding-document citations record references to the Treaty, Protocol and annexes without independently establishing legal authority. An association therefore remains separate from a paper's disposition. Agreement need not produce an instrument edge, and an edge need not establish agreement. Instrument citations likewise remain separate from operative actions. Table 1 summarises the distinctions.

In the working-paper cohort, 822 papers (51.5%) have some numbered-instrument association, including links to historical or later instruments. Across the full corpus, the 1,429 paper–instrument pairs comprise 834 same-meeting, 498 historical follow-up and 97 later associations. For special-meeting instruments without a regular ATCM ordinal, distinct adoption and submission years establish the temporal order. This resolves six links to Resolution 1 (2000) as historical follow-up. A follow-up paper concerns an instrument already adopted, and even a same-meeting association may record a shared subject rather than acceptance of a paper's recommendations. Disposition categories are paper-level classifications, so several can concern the same institutional product. Mixed or unclear outcomes account for 44 working papers (2.8%), withdrawals for seven (0.4%), and 77 (4.8%) were not identified in the report. The earlier successor search remains incomplete for the revised referral and deferral cohort.

Reporting conventions affect the visibility of these links. ATCM XIII (1985, paragraphs 62–64), for example, records draft recommendations, deliberation and adoption by delegation and subject without naming input papers by number, whereas ATCM XLIII (2021, paragraphs 124, 130 and 152) identifies WP16 and IP12 and records discussion, agreement on a work-plan update and adoption of Decision 5 (2021). Explicit identifiers already appear in some early reports, including ATCM XV (1989, paragraphs 40–43), so this is not a clean division between early and recent reporting, and the reconciled table retains eight early paper–instrument pairs, two from ATCM XI and six from ATCM XV.

### Recovering and checking relationships

Source PDFs were converted to Markdown with Marker/Surya, using model-assisted repaired text where available. Agenda items and early-report topic sections supplied the reading units. The parser retained original source spans and separate ATCM and CEP identities, reconciled extracted units against the report's agenda inventory, and flagged unresolved boundaries and unassigned discussion. In each structured interview, we supplied the full selected unit to Qwen3.8-27B and asked sequential questions about papers, instruments, relationships and evidence. The model was instructed to use only the supplied text, quote its support and preserve uncertainty. A separate Codex reading examined instrument texts and listed operative actions and founding-document citations. Model settings, the question sequence and coverage checks are documented in the supplementary methods (Supplementary Methods).

We resolved identifiers against the inventories and consolidated duplicate observations, retaining their provenance. Paper associations use a stronger tier for proposal or contribution labels and a weaker tier for shared-subject labels. Both enter the main descriptive summaries, with restricted comparisons reported separately. Associations run from paper to instrument. Timing compares submission with instrument adoption and classifies each association as same-meeting, historical follow-up or later. These categories do not establish when a proposal was accepted. Site-title matches supplement relationships where plans were discussed by site and adopted in groups, but retain their inferred status. Early reports often lack paper numbers, and the reconciled table retains only eight early-report pairs. Paper connectivity across the full period is consequently sensitive to coverage.

The analysis uses the 15 October 2026 extraction with source-reviewed operative and documentary overlays assembled on 20 October and a documentary source refresh on 25 October. It combines the complete 1,640-input Qwen rerun, model-based source review of 1,085 previously unlinked working papers, and accepted observations from an incomplete codebook-1.2 repair. Raw instrument-relation and founding-citation records are preserved separately, and the current analysis adds the reviewed documentary, named-object and founding-citation overlays. The operative action layer was rebuilt from the instrument readings. Source hashes, inventory resolution and quotation checks provide traceability, while repeated model readings measure consistency. None supplies an independent estimate of semantic accuracy.

A companion report corpus prepared on 25 October 2026 provides a later transcription layer, including a corrected 40-page ATCM IV report. It preserves source PDFs and raw text alongside deterministic cleaning and separate table exports. A bounded rereading of the ten restored Recommendations recovered 24 explicit numbered-reference pairs. One was already present and received additional source evidence, and 23 enter as new documentary pairs, including 19 endpoints of a single inclusive range. This source-specific overlay augments the earlier readings. It does not recode references as operative actions or replace the remaining historical extraction inputs.

## Paper outcomes

We classified outcomes for the 1,595 working papers of ATCMs 16–47. The categories distinguish instrument adoption, agreement or endorsement without an instrument, noting, referral, deferral, withdrawal, collective non-agreement, mixed or unclear outcomes, and papers not identified in the report. Collective non-agreement requires a recorded institutional outcome rather than an objection by an individual participant. We reconciled saved dispositions with source-review decisions and removed the rule that assigned adoption from a same-meeting instrument association. A fresh source reading assessed 442 papers with missing or disputed dispositions, using the relevant ATCM and CEP discussion and adopted texts. Adoption required evidence linking the paper's proposal to an adopted numbered instrument. Source assessments retain recommendation-level outcomes and distinguish committee endorsement from ATCM adoption. The final table contains 522 adoption dispositions, 425 agreements or endorsements without instrument adoption, 44 mixed or unclear outcomes and 77 papers not identified in the report, alongside the other recorded categories. These are source-assessed classifications, not calibrated adoption rates or evidence that every recommendation in a paper was accepted.

## Operative actions and maintenance

The recorded layer mixed operative provisions with historical recitals. We reconstructed operative relations from the reader's operative list and quotations, checking identities and current actors against adopted texts. Targeted source reviews checked narrative-only claims, action labels, retirement lists and recurring series. The November audit retained the resulting 531-pair candidate for comparison and added a separate source reconciliation. This removed one report-narrative succession claim from the operative graph and added 28 pairs, including component amendments and cessation clauses whose targets were explicitly identified within the acting text. Each change is recorded with its source passage as an overlay on the preserved extraction.

The reconciled operative graph contains 558 unique pairs, comprising 500 replacements or retirements, 56 amendments and two requests that Parties approve an instrument. Replacement includes changes to a plan, annex or language version, not necessarily cessation of the whole instrument. A clause covering prior plans contributes an edge when the same instrument explicitly identifies the affected plan and its numbered adopter. An outside registry association alone does not establish that relation. Seven bulk status-review Decisions supply 129 pairs, separately coded as spent, no longer current or withdrawal of never-effective instruments. We report the full graph and the graph without these bulk reviews. Requests do not establish approval. One amendment updates administrative information fields. Five literal targets with source identity conflicts remain flagged and are excluded in a separate sensitivity analysis. These model-assisted checks are distinct from independent human validation. The frozen recital-inclusive layer retains 1,893 action records after its established chronological exclusions and provides a comparison of relationship definitions.

To compare the model-assisted reconstruction with a rule-based reading, we also used the Secretariat's status and memo fields and fixed rules over adopted instrument text. The rules were frozen before comparison and recovered 555 pairs, including 515 of the 558 reconciled operative pairs (92.3%). The rule-based graph also concentrates revision within instrument forms and recurring plan histories. Additional pairs often come from status records that describe supersession without an explicit operative provision in the adopted text, as with the consolidated historic-site list of 2003. Under that broader reading, 12 rather than 21 of the 24 other Measures adopted in 1995–2012 remain untouched within twelve years. Pairs missed by the rules include entries in unparsed lists and annexes omitted from database pages. We use these differences to compare sensitivity to sources and definitions, while agreement between readings of the same sources does not establish accuracy.

Let $I$ be the numbered instruments and $E_{\text{op}}$ the directed pairs with a recovered operative relation retained for analysis. A pair $(i, j) \in E_{\text{op}}$ runs from acting instrument $i$ to target instrument $j$, dated no later than $i$. Action labels and evidence remain in the underlying records. Set $m_i = 1$ when $i$ has at least one such outgoing relation, and $m_i = 0$ otherwise. For instruments $I_B$ adopted in period $B$, the *maintenance share* is

$$M_B = \frac{\sum_{i \in I_B} m_i}{|I_B|}.$$

Each acting instrument counts once, even if it affects several predecessors. The criterion is inclusive, covering approval requests, and measures a relation between instruments rather than completed legal change. An instrument can act on a predecessor and introduce new provisions, and the absence of a recovered predecessor does not identify a new arrangement (predecessors may be unnumbered, outside the corpus or missed), so we do not estimate a creation share. We report maintenance for all instruments and after excluding site plans, historic-site listings and protected-area system rules, classified from database subject titles.

## Lineages, observation windows and agenda routes

For instrument histories, the cleaned operative graph uses $I$ as its nodes and $E_{\text{op}}$ as its edges. Edges run from the acting instrument to its target. Chronological succession displays reverse this direction and retain same-year relations, while the example arrows in Figure 4 retain actor-to-target direction. Action times refer to adoption, not entry into force or implementation. A lineage is a connected component containing at least two instruments when direction is ignored. Instruments without recovered action edges remain isolates. Lineages can join several sites through a common system-level instrument and need not describe a single policy object. Chain length counts instruments on the longest directed path. Graph checks and exclusions are specified in the supplementary methods.

To locate connections within and across decision streams, we count unique operative pairs by the instrument forms at their endpoints and by whether both, one or neither endpoint belongs to the title-based site category. Site-related instruments include management plans, historic-site listings and protected-area system rules. These counts measure the concentration of recovered links, rather than graph density normalised by the number of possible pairs. A two-instrument lineage is a pair, and a longer directed chain has at most one incoming and one outgoing edge at every node. Other components contain branching or conver-gence. Undirected cycles indicate alternative connecting paths. Their number of independent cycles is the number of edges minus nodes plus components. A Recommendation's participation in a connection between forms is distinct from its being an articulation point, whose removal disconnects its component when direction is ignored.

For comparison with operative histories, we also retain general instrument references from the recorded layer, apply the same endpoint and chronological filters, and collapse repeated endpoint pairs. The reviewed documentary graph includes recovered direct references, uniquely resolved references to versioned annexes and every known operative pair. It contains 2,782 distinct pairs from 3,379 chronologically retained records, and 20 further endpoint-eligible records fail the chronological filter. The form matrix retains all 3,399 records with eligible endpoints, including repeated observations, and therefore has a different denominator from the graph. Four citations nested inside reprinted older texts are retained only in sensitivity analysis.

We examined apparent isolation through a targeted, Codex-assisted source review of all 29 instruments without incoming or outgoing documentary links. Reading adopted texts, attached guidance and later report passages recovered three explicit references. Checking the remaining 26 identified a reference to the versioned 1997 Rules annex. These four links enter the documentary graph. References to named codes, websites, reports and declarations were also annotated against separate object identities. Six instrument-pair projections through these objects form an expanded sensitivity view, with their inferred status and unresolved editions retained. Shared subject matter or similar wording alone did not establish a reference. Source passages, hashes and resolution decisions accompany the annotations. This targeted curation and the later automated audit remain model-assisted source assessments.

We followed instruments adopted through 2014 to the end of the record in 2025. Actions in the same year count as one event, and the plotted pathway stops at the first replacement or retirement under the inclusive classification. This is an analytic stopping rule. Replacing a component or recording an instrument as spent does not establish the end of the whole instrument's legal life. An instrument without a recovered later action is censored at the observation endpoint and is not assumed to remain unchanged indefinitely. Kaplan–Meier summaries and fixed five-, ten- and twelve-year windows address unequal follow-up. The supplementary timing comparison for the 1995–2012 cohort uses a twelve-year window. The plotted Wilson intervals describe uncertainty in binomial proportions under that model. They do not quantify extraction error, and dependence among successive instruments limits an independent-trial interpretation.

The agenda-route comparison uses working papers of ATCM 22–47 assigned exclusively to CEP or ATCM agenda items and identified in the report. We attributed instruments to an agenda route when all their linked working papers with same-meeting or later associations belonged to that route. The instrument cohort covers adoption in 1998–2014 and follow-up through 2025. We compared paper outcome categories, subsequent operative relations and replacement instruments associated with new papers of the same route. Paper outcome comparisons use the reconciled submission-meeting dispositions, while instrument routes are assigned from the separately qualified association table. Sensitivity comparisons restrict association strength, timing, site-title inference and adoption period, or exclude site rules. These are descriptive comparisons. Topic, form and route overlap, and route attribution does not identify a committee effect.

Additional structural counts and sensitivity comparisons are reported in the supplementary results (Supplementary Results).

## Conditional network comparisons

To assess whether link concentration exceeds the available opportunities, we compared the observed graphs with directed target-swap networks preserving each instrument's incoming and outgoing link counts, adoption years and labels. Targets could not postdate actors, and self, duplicate and reciprocal pairs were excluded. The primary comparison also preserved each actor's target-age bands (0, 1–4, 5–9, 10–19 and 20 or more years). Further comparisons preserved target-form or title-class patterns. Title classes are inherited instrument-function descriptors, not independent policy-topic coding.

Twelve ensembles used two seeded chains of 500 draws each, after 200 times the edge count in burn-in proposals and with 20 times the edge count between draws. Holding times were retained. We report plus-one Monte Carlo tails and Benjamini–Hochberg adjustment for eight primary form/function comparisons. Null ranges describe conditional network variation, not extraction-error intervals. Degree preservation fixes isolates and mean lag, so we do not test those quantities as unexpected outcomes. The November rerun retained the earlier design and seeds, with the reconciled input tables frozen before sampling.

## Automated source audit

We audited frozen relationship tables using fresh Codex sessions, with source inventories completed before candidate links were revealed. The audit sampled 400 documentary pairs and 400 paper–instrument associations and examined all 531 pairs in the operative candidate. Separate source samples assessed omissions and document coverage. Readers used report text or original PDF images and retained evidence quotations. The frozen audit records identify the model as `gpt-6.1-sol`. All readers used the same model family, so their agreement measures reproducibility rather than independent validation of accuracy.

Source review identified an ambiguity between the stored replacement label and the broader definition that includes retirement, administrative status review and changes to instrument components. We retained the original audit and repeated the operative assessment after stating the intended definition explicitly and restoring missing source context. Nominal Krippendorff's $\alpha$ was 0.952, with a source-cluster bootstrap interval of 0.832–1.000. We checked disagreements against the adopted texts and applied corrections as versioned overlays. Unresolved source coverage prevents a whole-corpus recall estimate. The supplementary methods report the original audit, subsequent reconciliation and their limitations (Supplementary source audit).

Codex also assisted manuscript drafting and revision, using the GPT-6 family for the final editorial session. This editorial assistance was separate from the structured extraction and source-audit protocols.

## Research materials and declarations

**Acknowledgements.** I thank Zach and Larissa for helpful comments on an earlier draft of this work.

**Data and code availability.** The meeting reports, submitted-paper records and instrument texts are publicly accessible through the Antarctic Treaty Secretariat (Antarctic Treaty Secretariat, n.d.-d). The accompanying archive of research materials contains the reconciled relationship tables, analysis code, figure inputs and source-review evidence, together with version and checksum manifests. The October extraction and overlays remain preserved alongside the November reconciliation. Source manifests distinguish original reports from corrected or additionally retrieved documents. The machine-readable report release and the derived decision-network dataset remain separate products. The derived data and code will be deposited in a public repository by publication, with the verified accessions added to the published article.

**Ethics.** The study analyses published institutional documents and involved no recruitment of human participants.

**Author contributions.** Casper van Elteren developed the research question and methods, performed the analysis and wrote the manuscript.

**Funding.** This work was supported by the Australian Research Council Special Research Initiative Securing Antarctica's Environmental Future (SR200100005).

**Competing interests.** The author declares no competing interests.

**AI assistance.** Language models assisted documentary extraction, source review, manuscript drafting and editorial revision. We distinguish model-generated checks from independent human validation throughout. The methods describe the models and supplied evidence, and the retained responses document their role in the analysis.

# Supplementary Information

## Supplementary results

Excluding the update and both requests leaves 555 pairs, with the largest component unchanged at 330 and 249 isolates. Twelve instruments adopted through 2014 receive one action without replacement by the endpoint, nine have actions in two distinct years and three in three or more. Actions are grouped into events by calendar year. On the operative layer, 176 of 221 site rules (80%) receive an action and 34 of those (19%) receive another. Outside the site group, the counts are 139 of 322 (43%) and nine of those 139 (6%). Other components contain 25 and 22 instruments, and 34 lineages of two to four instruments contain 92 instruments. Four components contain alternative undirected paths, 49 of the graph's 55 independent undirected cycles occur in the largest component, and 11 Recommendations are articulation points. The longest chain ending at a Measure contains seven instruments, the longest path of Measures only six and the longest Recommendation chain three. Chain lengths count instruments, not the depth of revision to a text. In the full operative graph, 334 of 558 pairs connect two site-related instruments, 43 cross the title-based site boundary and 181 join instruments outside the site group. Without bulk reviews, the corresponding counts are 334, two and 93 of 429. The 30% within-form benchmark uses the observed acting and affected form shares and does not account for time or subject. Excluding the five source-conflicted operative targets leaves 553 pairs and reduces the largest component to 328, with 244 isolates. Excluding the 13 qualified documentary pairs leaves 2,769 pairs, while the largest component and isolate count remain 724 and 25.

The frozen recital-inclusive action layer has a largest component of 338 instruments and 272 isolates, and maintenance shares of 12%, 18% and 73% across the three periods (5%, 6% and 48% without site rules). These sensitivity comparisons between relationship definitions do not give bounds on the true shares. Adding six separately labelled connections through named institutional objects and codes to the documentary network increases its largest component to 727 and leaves 22 isolates. For agenda routes, two CEP-associated instruments (1%) are first amended and 27 (19%) have no recovered action, against four (5%) and 49 (65%) for the plenary. Route comparisons are inclusive histories of route-associated instruments, and return links include shared-subject associations and bulk retirement.

The matched-network comparisons support concentration beyond source activity, target popularity, chronology and coarse reference age. All eight primary form/function shares exceed all 1,000 draws of their respective nulls (plus-one upper-tail and adjusted values approximately 0.001). This is a conditional network comparison, not a causal or accuracy estimate. With form patterns held fixed, the non-bulk within-function excess remains 20.2 percentage points, and with function patterns fixed, the within-form excess is 3.3 points (Figure S10). All monitored non-invariant statistics passed the recorded R-hat and effective-sample-size thresholds, although finite-chain diagnostics do not prove exhaustive exploration.

Direct references have an 11-year median age and 53.6% reach back more than ten years. This does not establish a general ten-year attention limit. The opportunity comparisons distinguish that reference-age distribution from the concentration of site-plan maintenance at five/six years. For Measures with a complete twelve-year window, site plans receive 65 first returns in 138 instrument-years at ages five/six, against 28 in 516 at other ages. These descriptive rates use changing risk sets and do not identify a causal effect of the review rule.

## Extended methods

**Extraction and reconciliation provenance.** The frozen extraction combines the instrument readings and the complete October agenda-item rerun. Paper associations incorporate the earlier source review of previously unlinked working papers and accepted layers from the incomplete codebook-1.2 repair. The raw outputs and the 20 and 25 October overlays remain unchanged. The November audit and subsequent source reconciliation supply separate additions, exclusions, relation qualifications and working-paper dispositions. The reported graphs and figures use those reconciled tables. Source hashes distinguish the historical reports from restored or additionally retrieved source material, and the machine-readable change ledgers preserve the reason for each revision. Model-assisted review does not establish independent human accuracy.

### Corpus and document identities

We reconstructed documentary relationships from the English-language final reports of Antarctic Treaty Consultative Meetings (ATCMs) 1–47, covering 1961–2025. The reports contain meeting discussion and the texts of adopted Recommendations, Measures, Decisions and Resolutions. Source PDFs were converted to Markdown with Marker/Surya, and text repaired with model assistance was used where available. We retained the selected report version and source identifiers so that extracted relationships could be traced back to their report text. The corpus represents the published meeting record and does not cover every negotiation or intersessional exchange.

The later companion corpus contains 47 regular English-language report volumes, 14,795 source pages and 4,498,483 whitespace-delimited tokens. Its cleaned texts start from raw extraction and retain a replayable edit journal. The corrected ATCM IV source restores Recommendations IV-13 and IV-20–IV-28, whose original texts were absent from the earlier instrument reading even though their identities and some later incoming references were retained. The 25 October overlay incorporates the recovered explicit numbered references and retains their exact source spans and hashes. The refresh is confined to documentary links, with operative and founding-citation data retaining the preceding review and its unresolved coverage of the restored texts. The five special-meeting instruments in the graph retain separate sources outside this regular-report package.

A table supplement provides 2,692 recognised Markdown table blocks and 2,113 native-PDF candidates from the cleaned reports. These overlapping views retain cell strings, source evidence and review flags. They are not a census of distinct tables and have not been human validated. They were not inputs to the reported graph, and successful export does not establish correct cell boundaries or reading order.

Separate inventories supplied the documents against which the extraction was checked. The Antarctic Treaty Secretariat's instrument database supplied instrument types, adoption dates, numbers and subject titles. Its records were fetched by meeting, with retrieved counts checked against the totals returned by the database. The Antarctic Database supplied submitted-paper identifiers, titles, submitters and agenda assignments. Working and information papers were consolidated by meeting, paper type and number, so revisions and language editions did not become separate papers. Secretariat and background papers named in the reports could also enter the extracted relationships, although they were outside the working- and information-paper inventory used for coverage summaries.

Instrument identity retained the type, number and adoption year, or the historical meeting–sequence designation for Recommendations. Thus Measure 1 (1995) and Measure 1 (1997) remained distinct instruments. Numbered special-meeting instruments were retained in the inventory, with their available report texts read separately. Declarations, statements and other entries outside the four numbered instrument types were recorded separately and excluded from the rule analysis.

### Report structure and extraction units

The unit of the structured interview was an agenda item or, in early reports, a recovered topic section (Figure S2 illustrates how report structure, paper identities and adoption anchors be-come documentary links). We located discussion using agenda headings, adopted-agenda inventories and paragraph markers. Recognition accommodated topic headings, titles followed by an agenda-item suffix, item numbers and titles split across lines, and modern inline item headings. Embedded Committee for Environmental Protection (CEP) reports were identified separately from the ATCM discussion that summarised their advice, retaining the CEP's own agenda and paragraph numbering. Adopted instrument texts were read in a separate extraction stage.

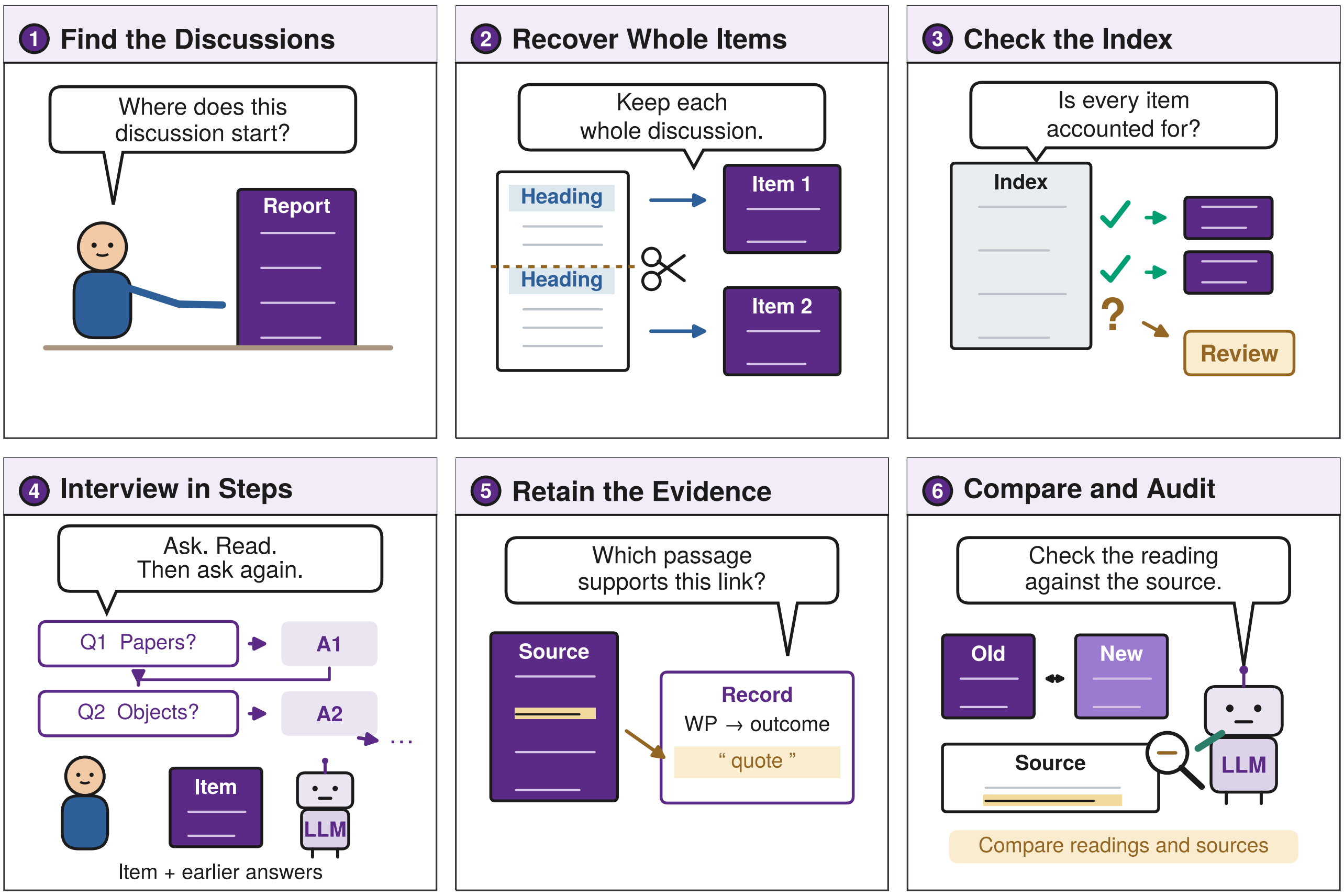

**Figure S1: From reports to traceable records.** Heading-bounded discussions are checked against the agenda index before model reading, and unresolved scopes remain available for review. The interview proceeds one question per turn, carrying the source item and earlier answers forward. Candidate relationships retain supporting quotations. Comparing paired reruns measures stability, while fresh model readings assess source support and record unresolved coverage. The panels illustrate the workflow rather than audited performance.

After parsing, we cross-referenced the source-addressed extraction against the report's own agenda index or adopted-agenda list. Every index entry, including identified subitems, required one recorded outcome, which was either extracted discussion text, no separate text as documented by source evidence, or unresolved attribution with a reason. An independent lexical scan of the located index regions checked for entries omitted from this inventory. Extracted outcomes had to point to their exact discussion spans, and a missing parser match alone could not establish that no discussion existed. Numbered discussion paragraphs also had to be assigned to an item or explicitly recorded as unassigned. These structural checks preceded preparation of the additional model interviews.

Initial runs used heading-based agenda cuts. Subsequent runs used a source-addressed extractor that retained original line spans, unnumbered continuations and separate ATCM and CEP identities. A coverage ledger located previously processed text in the selected reports and identified new discussion or uncovered fragments. Existing completed interviews were reused, and additional interviews processed the recovered gaps. Source hashes, span replay and non-overlap checks assessed text selection and provenance. These checks establish which text the reader received, but do not establish whether its relationship judgments were correct. Unresolved boundaries and unassigned discussion remained recorded as coverage limitations.

The current parser treats printed headings and item transitions as the primary boundary evidence. Format-specific rules accommodate the different publication structures, especially in older reports, while substantive source wording is retained. Index matching is a separate check on identity and coverage. A clearly labelled discussion block can therefore be retained even when its numbered agenda identity is uncertain. Parent headings without their own prose remain structural records, while discussion under their children is assigned to the child units. Shared or resumed discussion can have several original fragments, whose order and joining separators are retained in the manifest.

An automated check tests source coordinates, exact replay of source passages, overlaps between items, section membership, accounting for index entries and coverage of numbered paragraphs. It also inspects unowned text between successive source owners. Blank lines, page headers and table furniture do not constitute missing discussion. An unowned numbered paragraph triggers a flag regardless of length, and other intervening prose is flagged when it exceeds three lines or 80 characters. Leading and trailing unassigned blocks are recorded separately for inspection. These flags identify coverage or attribution questions and do not measure the error rate of the model's answers. Older heading variants and unresolved subitem identities remain targets for source spot checks.

In Algorithm 1, a source unit $u$ retains its original line spans $\Sigma_u$, verbatim text $x_u$, and provenance $\pi_u$, including the selected report and text hashes. The coverage ledger $\mathcal{L}$ records the outcome for every observed agenda identity, while $\mathcal{E}$ retains unresolved identities and unassigned source evidence. Candidate records $\mathcal{D}$ carry diagnostic flags rather than an automatic claim of source support.

**Input:** selected reports $\mathcal{R}$; codebook $\mathcal{Q}$; frozen comparison records $\mathcal{B}$.
**Output:** candidate records $\mathcal{D}$; coverage ledger $\mathcal{L}$; exceptions $\mathcal{E}$; paired comparisons $\mathcal{C}$.

$\mathcal{D} \leftarrow \varnothing$; $\mathcal{L} \leftarrow \varnothing$; $\mathcal{E} \leftarrow \varnothing$

**for each** $r \in \mathcal{R}$ **do**

$(X_r, \pi_r) \leftarrow \mathrm{FreezeSource}(r)$

$G_r \leftarrow \mathrm{LocateRegions}(X_r)$

$(A_r, U_r) \leftarrow \mathrm{ParseAgenda}(X_r, G_r)$

$(L_r, E_r) \leftarrow \mathrm{ReconcileIndex}(A_r, U_r, X_r)$

**if** structural invariants fail **then** record source failure and **halt**

$\mathcal{L} \leftarrow \mathcal{L} \cup L_r$; $\mathcal{E} \leftarrow \mathcal{E} \cup E_r$

$U_r \leftarrow U_r \cup \mathrm{LabelledUnindexed}(X_r, G_r) \cup \mathrm{ReviewContexts}(X_r, G_r)$

**for each** $u \in U_r$ **do**

$\Sigma_u \leftarrow \mathrm{OriginalSpans}(u)$

$x_u \leftarrow \mathrm{Slice}(X_r, \Sigma_u)$; $\pi_u \leftarrow (\pi_r, \Sigma_u, \mathrm{SHA256}(x_u))$

$H_u \leftarrow \mathrm{InitConversation}(x_u, \mathcal{Q})$

**for each** $q \in \mathrm{Questions}(\mathcal{Q})$ **in order do**

$a \leftarrow \mathrm{LLM}(H_u, q)$; $H_u \leftarrow \mathrm{Append}(H_u, q, a)$

**end for**

**if** $\mathrm{OutcomeRequired}(u)$ **then**

$a \leftarrow \mathrm{LLM}(H_u, q_O)$; $H_u \leftarrow \mathrm{Append}(H_u, q_O, a)$

**end if**

$d_u \leftarrow \mathrm{ParseResponses}(H_u)$

**if** no response layer yields a usable record **then** persist failure and **continue**

$v_u \leftarrow \mathrm{CheckSchemaIDsQuotes}(d_u, x_u)$

$\mathcal{D} \leftarrow \mathcal{D} \cup \{(d_u, \pi_u, v_u)\}$; persist the record and conversation

**end for**

**end for**

$\mathcal{C} \leftarrow \mathrm{ComparePaired}(\mathcal{D}, \mathcal{B})$

**return** $(\mathcal{D}, \mathcal{L}, \mathcal{E}, \mathcal{C})$

**Algorithm 1: Complete-context documentary extraction.** $G_r$ denotes source regions, $A_r$ the observed agenda inventory, and $U_r$ the recovered units. $H_u$ retains the instructions, source text and previous answers. The ordered questions are Q1–Q6 followed by I1–I4, with $q_O$ the optional OUTCOMES question. The record checks $v_u$ remain diagnostic flags, and review contexts remain distinguishable from numbered agenda items. Paired comparison measures reading stability, and the source audit separately assesses claims against the supplied reports.

**Table 2: The structured-interview sequence.** Each question was a separate turn in the same conversation. Q1–Q6 extracted paper–instrument associations, and I1–I4 extended the interview to relations among all mentioned instruments. The versioned codebook retains the full wording and output schemas of the questions summarised here.

| Question | Task | Requested information |
|---|---|---|
| Q1 | Identify papers | Identifiers as written and who introduced each paper. |
| Q2 | Identify objects | Named sites, protected areas, provisions, guidelines or other discussion objects. |
| Q3 | Identify adoptions | Instruments said to have been adopted, the adopting body and their objects or subjects. |
| Q4 | Examine paper–instrument pairs | Pairs with a textual basis, supporting words, explicit statements or a step-by-step inference, and papers without a candidate link. |
| Q5 | Test inferred relationships | A label, justification, unstated assumptions and the additional evidence needed to settle each inference. |
| Q6 | Record paper relationships | JSON entities and relations, with stated/inferred/undetermined status, quotations and assumptions. |
| I1 | Identify all instruments | Every mentioned instrument, including earlier instruments and unnamed groups, with paragraph locations. |
| I2 | Locate inter-instrument relations | Exact wording, stated or inferred support, and acting and affected endpoints. |
| I3 | Label and test relations | The action or reference label, assumptions and evidence needed to resolve uncertainty. |
| I4 | Record instrument relationships | JSON relations with exact identifiers, status, a contiguous quotation and assumptions. Unnamed targets remain unresolved. |

## A structured interview with the model

We used a large language model as a reader in a fixed, sequential interview. Each agenda item began a separate conversation. The model received instructions to work only from the supplied text, use no tools or outside knowledge, quote evidence, and state explicitly when a relationship could not be determined. Questions were sent one at a time, with the source text and previous answers remaining in the conversation. This separated the identification of documents and discussion objects from the subsequent judgments about their relationships.

The first six questions identified papers, named objects and adopted instruments, examined possible paper–instrument pairs, tested the assumptions behind inferred links, and produced a structured JSON record (Table 2). The record retained document identifiers as written, relation labels, evidence quotations, assumptions and a status distinguishing what the text stated from what the reader inferred. The paper–instrument vocabulary comprised proposal followed by adoption, documented contribution, shared subject only, unrelated and undetermined. Proposal and contribution labels formed the stronger association tier, and shared subject the weaker. These tiers describe the assigned relationship, rather than calibrated confidence or demonstrated causal influence.

Codebook versions 1.0.2 and 1.1.0 used the same Q1–Q6 questions. Version 1.1.0 added four questions about all mentioned instruments, including those not adopted at the meeting. This extension distinguished replacement, incorporation of provisions, amendment, revocation, designation as no longer current, and reference without a specified action. The setup occupied two turns, followed by six paper questions and four instrument questions in the twelve-turn protocol. Earlier eight-turn interviews were extended by replaying their stored conversations and generating only the four additional answers, preserving their original paper judgments.

We ran Qwen3.8-27B through vLLM on the NCI Gadi supercomputer using NVIDIA H200 GPUs. Saved engine configurations record FP8 quantization with a BF16 runtime dtype. The saved configurations do not identify the exact model-weight revision. Production sampling used temperature 1.0, top-p 0.95 and top-k 20. Prompts, answers, codebook versions, source-text hashes, generation metadata and parsing outcomes were retained. Because generation was stochastic, a repeated interview could return different candidates or labels.

## Matching submissions in early reports

For ATCM 1–15, discussion often described a submission by delegation and topic without its paper number. We therefore supplied a separate reader with the recovered section and closed lists of that meeting's database papers and adopted Recommendations. The reader matched described submissions to candidate titles and linked them to Recommendations named in the section or described as adopting the draft. It used the same proposal, contribution and shared-subject labels, returned evidence quotations, and recorded unmatched submissions rather than inventing identifiers.

We requested three generations per section and retained a paper–Recommendation pair when at least two reported the same endpoints. Its label was the most frequent label among the generations that reported the pair. Where section-level candidates produced duplicate pairs, the stronger retained label took precedence. These links entered the analysis as inferred early-report matches with separate provenance. Agreement between generations was a stability criterion, not an accuracy estimate. An early paper without a recovered match could not reliably be distinguished from an unnumbered submission discussed in the report.

## Instrument texts and founding provisions

A separate Codex model reading examined the adopted instrument texts, usually in Part II of the final report or an annex in earlier reports. An official checklist required one record per instrument, including an explicit missing-text record when its text could not be found. The reader recorded cited governing documents and articles, recalled instruments and supporting quotations, and listed separately the operative actions of each instrument on other instruments, with the action (revoke, replace, amend, approve or adopt), the target as written and a quotation. It read from the instrument heading through its operative paragraphs. Long attached management plans, guidelines and financial annexes were excluded from this stage.

The instrument-text reading distinguished designation, adoption, amendment, supersession, revocation and reference. These records were merged with the interview's instrument relations while retaining their different source vocabularies. In the recorded layer, supersession, revocation, incorporation of provisions and designation as no longer current were grouped as replacement or retirement. Amendment remained separate, and adoption and designation were grouped as carrying an instrument forward. References without an assigned action were retained in the documentary tables but excluded from action pathways.

Founding-document records retained the cited document, article when specified, citation context and quotation. We summarised these citations using the Treaty, Environmental Protocol, its annexes and other governing documents. For the stream displays, each instrument was assigned to its deepest recorded category, with Annex V distinguished from the other annexes, followed by the Protocol, Treaty and other documents. An instrument without a recovered citation had a separate category. This assignment summarises recorded citations and does not independently establish an instrument's legal basis or validity.

## Identity and evidence checks

We checked returned records for valid JSON and conformity to the schema and vocabulary. Parsing fallbacks handled escaped JSON, extraneous text, malformed closure or complete entries recoverable from truncated answers. The parsing route and failures remained recorded, so a salvaged partial record was distinguishable from a complete response. Paper and instrument identifiers were then resolved against the inventories. An untyped paper received a type only when its meeting-specific number uniquely matched a database type. A modern instrument without a year could receive the year of an explicitly identified meeting, but annual numbers alone were never used to merge instruments.

At extraction, returned quotations were checked against the supplied item text after formatting normalisation. The release assembly used a less stringent check. After lowercasing and normalising whitespace and Markdown markers, it tested the first 120 characters of each quotation segment separated by an ellipsis against the report. The stored quotation-verification flag therefore records this check of quotation prefixes and does not indicate full verbatim verification. A matching quotation also does not establish that it supports the assigned relationship.

In the original release, failed quotation checks, missing inventory identities, unresolved paper types and uncertain meeting assignments triggered a further Codex reading of the relevant agenda item. The reader returned a keep, fix or drop decision with a reason. Separate correction readings supplied candidate paper links for instruments whose discussion had not been processed. Retained uncertainties remained flagged, and dropped links and their reasons remained in a separate table. Newly processed text added resolved, previously absent positive pairs to the staged analysis dataset. Conflicting judgments and unresolved references were not automatically substituted for existing records. Early-report, new-run and site-title additions retained their staging status and were not treated as having undergone the original release's flagged-link review.

The final tables retained canonical and original identifiers, relation labels, status, evidence, extraction source and review information. Duplicate observations were consolidated according to their endpoint and relation keys, retaining source provenance and preferring stronger paper-association labels where the implementation required one label per pair. Model interviews, correction readings and review decisions were all machine-generated. Inventory membership and source anchoring were implementation checks.

## Complete-context rerun and consistency comparison

The full rerun replaced the incremental selection of old interviews and gap fragments with a fresh reading of every current agenda unit. Its frozen manifest contains 1,637 ATCM and CEP units, including 72 parent-heading records without independent prose. Two labelled Messages blocks from ATCM 7 and 8 are retained separately without invented agenda numbers, and one cross-topic paragraph from ATCM 10 remains a review context. These three additional blocks bring the input manifest to 1,640 records. Nine agenda-index identities remain explicitly unresolved. The index inventory records those uncertainties alongside extracted outcomes and outcomes independently documented as having no text.

Each request embeds the complete selected unit, original fragment spans, source and text hashes, meeting and section, and a stable run label. No unit is cut at a character limit or divided to fit a shorter job. Runtime estimates from earlier completed interviews are used to batch whole units for requested walltimes of 45, 60, 90 or 120 minutes, with time reserved for model startup and a margin on estimated work. The prepared plan has 54 jobs, with 34, 13, six and one in those respective classes. Batching jobs affects scheduling without changing the model's per-item con-versation or the extraction boundaries. Output directories and submission receipts distinguish the rerun from earlier results.

For a comparison defined before inspecting the new answers, we froze previous usable Qwen3.8-27B/codebook-1.1.0 records whose supplied text covered one complete current item and whose Q6 quotation checks passed. The baseline contains 740 previous readings across 666 current units. Of these, 551 readings across 550 units have byte-identical input text, and 189 readings across 116 units cover the same complete source prose with differences in formatting or heading context. The identical-input group provides the stricter consistency comparison. The other group is reported separately because its supplied context changed. Instrument-relation comparisons also retain the previous chain-validation status. The completed automated comparison contains all 740 reading pairs, 551 with identical input and 189 with the same complete source prose. Entities, endpoints, labels, stated/inferred status and evidence wording are compared separately. Agreement measures reading stability, not accuracy. The main jobs yielded 1,631 finalised item artifacts, and three follow-up jobs supplied the remaining nine same-input Qwen readings. All 1,640 inputs now have finalised Qwen artifacts, and the earlier provisional independent-model fills have been superseded. Finalisation does not establish the validity of every structured layer. The current paper tables incorporate the complete main rerun, source review and accepted repair observations, while retaining the earlier instrument-relation and founding-citation layers.

## Model-based source review

The 20 October operative review targeted distinct failure modes rather than drawing a population audit sample. It screened 54 links supported only by narrative quotations, checked the bounded retirement lists, examined successive plan and nonsite series, and reviewed the 42 documentary isolates in the original graph. Accepted changes added 134 operative pairs, removed 13 further false targets and corrected one action class. Four earlier exclusions remain in the correction table. Ordinary documentary additions and three founding-citation corrections are stored as separate overlays. Source quotations, actor and target identities, action scope, held cases and before/after hashes accompany these changes. These model-assisted checks do not establish precision or recall for the whole candidate.

We conducted a model-based source review of all 1,085 working papers without a numbered-instrument association in the pre-audit candidate. Six independent model readers received frozen source packets without the earlier Qwen predictions. Packets contained complete agenda units, paper identity metadata and the official instrument inventory. Additional relevant ATCM, CEP and annex passages retained source line coordinates. A separate reading reviewed the proposed associations, accepting 358 claims on 314 papers and quarantining four claims. Further adjudication of 70 papers separated their submission-stage dispositions from later associations. Mechanical validation checked quotation presence and source coordinates, but neither these checks nor agreement between model readers establishes semantic accuracy. Previously linked papers outside this target set were not all reread. In the fixed cohort of 1,595 working papers of ATCM 16–47, the review raised the number with a numbered-instrument association from 510 (32.0%) to 819 (51.3%), and the subsequent integration and source adjudication to 823 (51.6%). The net increase of 313 combines association gains for 318 papers with five removed mappings. Three of the four post-review gains are explicit associations with historical instruments, and the fourth is a documented contribution to an adopted work-plan Decision. These associations do not by themselves establish adoption of each paper's recommendations.

The reviewed candidate informed codebook 1.2.0 and a selective repair of 393 flagged agenda units. The revision distinguishes instrument state, institutional body, unnumbered products, mixed recommendation-level outcomes and collective decisions from individual positions. Repair packets retain the complete original item plus explicitly sourced additional context selected during review. Structured final responses are checked for schema, identities, quotations and requested-paper coverage. This comparison evaluates the revised codebook, validation and supplied context together and does not isolate prompt effects or establish automatic context retrieval. The 20 repair jobs stopped after a queue safeguard treated three consecutive record-validation failures as a reason to halt. Eleven interviews completed, 175 failed and 207 were interrupted. We evaluated saved final responses from completed, failed and interrupted contexts separately. Of these, 42 Q6 records passed unchanged, and 20 further Q6 records passed after typesetting and explicit identifier-alias normalization without changing substantive labels or adding evidence. Thirteen I4 records and eight OUTCOMES responses passed their layer checks. An interview remained incomplete even when one of its layers passed validation. Novel associations and changed dispositions received an additional source reading, and conflicting unadjudicated predictions did not replace the source review. The comparison remains selected and incomplete and does not estimate corpus accuracy.

A further model reading adjudicated eight disputed outcome classifications and the returned association discrepancies using the original report and adopted instrument text. Two aggregate dispositions changed. ATCM22:WP 8 was classified as noted, and ATCM20:WP 15 as mixed, preserving acceptance of its voluntary management plan and conditional deferral of formal ASMA designation. The annex to Decision 5 (2021) explicitly names the SCAR survey report and its recommendations, resolving the earlier quarantine of ATCM43:WP 16 as a documented contribution, and its supporting IP 12 received a shared-object association. Neither instrument association establishes agreement on every substantive recommendation. Original model responses and earlier judgments remain preserved.

## Automated audit and source reconciliation

The November audit froze the existing tables, source files and sampling frames before reading. Three fresh Codex readers assessed 400 sampled documentary pairs, 400 sampled paper–instrument associations and the census of 531 operative pairs. Recall inventories were collected before candidate pairs were disclosed. Separate samples of 240 source units assessed documentary, operative and paper-association coverage. Text readers used the report transcription, while a visual reader used original PDF pages. All readers used the same model family, and a further model reading adjudicated disagreements. Original responses, source hashes and unresolved cases are preserved.

For the documentary sample, the original adjudication supported 390 of 400 pairs, rejected one and left nine unresolved. Subsequent examination of complete instrument texts rejected another pair. Restoring the corrected ATCM IV report and original plan attachments resolved five cases, leaving 395 supported, two rejected and three unresolved. This later result uses additional source material and is reported separately from the frozen audit. Review of 117 proposed documentary omissions verified 98 missing pairs. Other proposals relied on unnumbered objects, contained conflicting literal identifiers, duplicated an existing pair or cited the source itself. The accepted additions and rejected pairs enter the revised graph as separate overlays. Nine further literal references whose identifiers conflict with other source information are retained with flags, applying the same rule used for existing conflicted observations. The chronological graph still excludes references to later years and both directions of a reciprocal same-year pair. Verified source occurrences therefore need not produce an equal net increase in graph edges.

The paper-association audit supported the exact assigned relation for 177 of 400 pairs, rejected 84 and left 139 unresolved. These outcomes assess both endpoints and the claimed relation, so rejection need not mean that the documents are unrelated. Most rejected claims attributed contribution to papers that discussed an earlier instrument or shared its subject. We therefore retain source-backed associations separately from claims of adoption or contribution and no longer use a same-meeting association to establish a working paper's adoption outcome. The revised disposition tables require evidence of the recorded response at the submission meeting, and unresolved responses remain explicit. The 84 rejected association claims yielded 79 source-backed relabellings and five quarantined pairs. Review of 27 proposed omissions added 26 associations, including 24 shared-subject links and two proposal-to-adopted-instrument links, while one title-only candidate remained unresolved. The original 139 unresolved association assessments remain recorded and do not become supported through relabelling other cases.

The original operative assessment supported 386 pairs, rejected 130 and left 15 unresolved. Examination of the rejection reasons showed that the stored replacement label had been read more narrowly than the manuscript's definition, particularly for administrative retirement and changes to a plan or annex. Some readers also lacked the continuation of a retirement list or readable images of scanned pages. We retained those results and conducted a separate post hoc recheck of all 531 pairs. The revised instructions defined the inclusive relation explicitly and distinguished a documented operative provision from subsequent approval, entry into force or implementation. Source packets combined the complete official database text with relevant original report pages. Scanned pages were supplied as images, and retirement-list continuations were included. Responses were reused only where the request and image-page list were identical between the initial recheck and its source-context repair.

In the corrected-definition recheck, all readers supported 524 pairs and flagged discrepancies for seven. Source adjudication separated a report-narrative succession claim from the strict operative graph, retained literal target conflicts as flagged observations and documented discrepancies between report headings, annex histories and database text. The source-miss review supplied 19 additional operative pairs, including three literal source conflicts. Applying the same component-scope criterion to the earlier 16-pair blanket-clause sensitivity set verified nine further pairs, found three overlaps and retained four exclusions. These additions were source-reviewed separately from the frozen 531-pair census and are not part of its agreement coefficient. The model ratings were not changed after adjudication.

Nominal Krippendorff's $\alpha$ was 0.952 and raw pairwise agreement was 99.87%. Resampling the 352 acting-instrument clusters gave a 95% interval of 0.832–1.000 over 1,999 defined bootstrap replicates out of 2,000. This interval describes variation across sampled source clusters, not independent model draws or a probability that the graph is correct. The original nominal $\alpha$ values were 0.506 for documentary claims, 0.730 for paper-association claims and 0.431 for operative claims under the earlier instructions. The higher value after reconciliation reflects a changed definition and supplied context and cannot be interpreted as an independent accuracy gain.

Source completeness remains a limitation. The original documentary recall calculation covered 136 of the 240 sampled units and recovered 431 of 468 reference edges in that subset. The resulting 92.1% is conditional on those units and does not estimate whole-corpus recall. A report-wide readability rule also allowed an unreadable page elsewhere in a report to mark an otherwise accessible instrument incomplete. Reconciliation distinguishes such packet limitations from missing instrument text and unresolved relation scope, but does not certify every annex as complete. No whole-corpus recall estimate or independent human accuracy estimate is claimed. Quotations and source coordinates permit inspection of individual decisions, while sensitivity analyses show how qualified edges affect the reported structure.

## Adoption, follow-up and site matches

Paper links were classified by the meeting of submission relative to instrument adoption. Same-meeting links were termed adoption ties for timing, without establishing adoption of the paper. Links to an already adopted instrument were follow-up ties, and links to a subsequent instrument were later ties. Follow-up papers may discuss, apply or seek to revise an earlier rule, but cannot have proposed its original adoption. Summaries of papers associated with a rule used adoption and later ties, keeping follow-up and unknown-timing links separate. Both stronger and weaker association tiers entered the principal coverage and route summaries, with their labels retained for inspection.

Protected-area plans were frequently discussed by site and adopted in groups without a paper-to-Measure mapping. We supplemented the interview links by matching site namespaces (ASPA, ASMA, SSSI and Historic Sites and Monuments) and numbers in database paper titles and instrument subjects. A match joined documents at the same meeting, or within the next two meetings when neither endpoint had a same-meeting site match. Among later matches, the nearest meeting was retained. Added pairs were marked as inferred site-title matches and analysed separately in sensitivity comparisons. They did not require a report quotation or demonstrate that the titled paper supplied the adopted draft.

## Disposition of working papers

In the earlier supplementary classification, model readers classified working papers of ATCM 16–47 that were named in the report but had no adoption tie. This included papers linked only to an earlier rule. Each packet contained the database title, submitters and up to three passages naming the paper, including the following paragraph when a passage was short. Excerpts were bounded at 1,500 characters. Readers used only the supplied passages and returned a disposition, a supporting quotation, any named instrument or actor, and a confidence category.

The disposition categories were adopted instrument, agreement or endorsement without an instrument, noted, referred, withdrawn, deferred, not agreed, and unclear or absent. Where several applied, the instructions prioritised adoption, non-agreement, withdrawal, deferral, referral, endorsement and noting in that order. The collector stored category-validity and quotation-found flags but did not exclude records solely because those flags failed or confidence was low. Figure summaries used the collected categories. A quotation-found flag tested text presence, not whether the quoted passage established institutional agreement. The route machine grouped noting, referral, deferral, withdrawal and unclear outcomes as other outcomes, while the detailed machine retained the distinctions. Duplicate packets supplied a separate measure of reader agreement and did not provide a human validation set.

The full rerun added an OUTCOMES turn within complete-item conversations for the working papers covered by this supplementary classification. It distinguished collective action by the Meeting or Committee from support expressed by most or individual Parties. An explicit failure to reach consensus overrides majority support, and a draft Measure is distinguished from an adopted instrument. The reader was asked to retain distinct recommendation-level outcomes and mark insufficient context, including where CEP advice did not establish the ATCM disposition. The frozen manifest assigned 1,223 paper identities to 498 conversations, and 27 identities without a verified current item owner remained in the unresolved inventory. Complete context addressed the excerpting failure illustrated by ATCM44 WP 34, where the earlier packet retained the passage reporting majority support but its character limit excluded the later CEP paragraph recording no consensus. The earlier assembly combined full-item outcomes and source-review corrections but assigned 352 papers to adoption through a same-meeting instrument association. Of these, 128 conflicted with saved disposition labels. The November reconciliation removed that shortcut and resolved labels from the saved evidence and fresh source readings. All 442 papers selected for source completion received an assessment, including 343 without an eligible disposition in the initial reconciliation. The new assessment retained 44 mixed or unresolved outcomes. Numbered adoption required separate evidence of the paper, the collective adoption and the corresponding instrument text. The source record preserves the responsible institutional body and distinct recommendation outcomes, so CEP endorsement or an individual party's support cannot establish ATCM adoption. Conflicting claims without sufficient source evidence remain unresolved. The resulting aggregate categories are mutually exclusive for display, while their underlying com-ponent outcomes remain available for inspection. The agenda comparisons for papers use these dispositions directly and do not infer adoption from instrument links.

For deferred or referred papers, candidate successors were retrieved from the next eight meetings using TF–IDF similarity of title words and word pairs, with a 0.1 bonus for shared submitters. Up to three candidates with a combined score of at least 0.35 and their report passages were supplied to a reader, who judged whether the same proposal had returned and classified its later disposition. Title resemblance alone did not establish continuity, and an unmatched paper remained unresolved.

## Rule pathways, lineages and timing

The recorded actions came from the recalled-instrument field of the instrument readings, which mixes operative provisions with preamble recitals of what earlier instruments did. We therefore built an operative action layer. It combined the reader's separate list of operative actions, whose targets were resolved to ATS instruments when they named an instrument dated no later than the acting one, with recorded actions whose quotation had operative wording, such as 'is no longer current', 'be revoked' or 'shall replace', and no recital wording, such as a preamble participle or a historical relative clause. Following the November source reconciliation, the operative graph contains 558 directed pairs, comprising 500 replacements or retirements, 56 amendments and two requests for Parties to approve an earlier instrument. Replacement includes partial component changes and administrative retirement, which are not uniformly whole-instrument legal endpoints. Seven bulk status-review Decisions supply 129 of the pairs. The two approval requests are Resolution 1 (1998) concerning Recommendation XVI-10 and Resolution 7 (2014) concerning Measure 4 (2004), and neither establishes that approval occurred. The reported inclusive summaries retain both requests. One pair, Decision 1 (2019) acting on Measure 12 (2019), has endpoints adopted in the same year, so targets are dated no later than the acting instrument rather than necessarily in a preceding year. The recorded layer, including recitals, supplied a sensitivity comparison rather than a bound on the true count, and lineages, chain lengths and timing in the main figures use the operative layer. The action analysis used relations between resolved ATS instruments, excluding self-relations and references without an action. The stored direction ran from the acting instrument to the affected instrument. Succession displays reversed this direction while retaining same-year relations, whereas the example arrows in Figure 4 retain actor-to-target direction. The recorded layer contained 1,896 action records after endpoint filtering. Removing one backward-dated relation and both directions of a reciprocal same-year pair left 1,893 records. These exclusions apply to the recorded layer and are absent from the separately assembled 558-pair operative graph. Acyclicity was checked for the graphs used in path calculations. The complete documentary-reference table was retained separately.

## Targeted review of documentary isolation

The 20 October connectivity review began with the 29 isolates in the documentary graph, considering incoming and outgoing links together. We inspected their declared source scopes and screened all 47 local final reports for identifiers and named objects. The source review recovered Recommendation III-VIII's reference to Recommendation III-VII in Article XIII(2) of the Agreed Measures, Resolution 1 (2016)'s reference to Resolution 2 (1997) in its annexed EIA Guidelines, and Resolution 1 (2024)'s reference to Resolution 3 (2016) in its annexed ASPA plan-preparation guide. Among the remaining 26 cases, a second review resolved Resolution 5 (1998)'s reference to Rule 2 of the Revised Rules of Procedure (1997) to the annex adopted by Decision 1 (1997). These four source-supported additions retain their source and target identities, verbatim passages, line spans, source hashes and resolution method.

The review also retained 25 observations involving 20 named document or institutional-object identities, together with individual assessments of all 26 remaining cases. An instrument's explicit reference to an object was distinguished from the association between that object and the instrument that adopted, endorsed, established or commended the object. The expanded graph adds six instrument-pair projections. Decision 1 (1999) is linked to Resolution 5 (1998) through the ATCM Home Page, Resolution 5 (2018) to Resolution 2 (2017) through SCAR's subglacial code, and Resolutions 1 (2016), 1 (2019), 1 (2024) and 2 (2025) to Resolution 3 (2015) through the Antarctic Environments Portal. The code title identifies its family but does not establish the exact edition, and the projected link to the 2017 endorsement retains that uncertainty. The Portal references concern the resource and do not explicitly cite its 2015 commending Resolution. External declarations, report editions, procedural rules and report paragraphs remain distinct object annotations unless a numbered source can be resolved. Similar mining-ban wording, generic references to current Resolutions and later implementation reports do not create direct instrument citations or operative actions.

The corrected ATCM IV report restores IV-13 and IV-20–IV-28, whose original texts the saved reading marked missing. The 25 October review checked their boundaries against all 28 Recommendations and the following explanatory statement. IV-20 cites IV-1–IV-19 inclusive, IV-21 cites III-XI, IV-22 cites I-IV, and IV-23 cites I-VI, III-I and III-II. The IV-22 reference was already in the reviewed layer, so it is reanchored without a duplicate observation. The other 23 pairs enter a separate overlay. The inclusive range yields 19 endpoint pairs from one source occurrence. Unnumbered Agreed Measures, unspecified effective Recommendations and external meeting documents are not forced into numbered targets. None of these citations enters the operative graph merely because it appears in an operative paragraph.

The documentary graph has 2,782 pairs, a largest component of 724 and 25 isolates. The expanded view has 2,788 pairs, a largest component of 727 and 22 isolates. Object projections are excluded from the main reference-form matrix and from every operative analysis. The reconciled operative graph contains 558 pairs. The source review and curation used Codex assistance and do not provide independent human accuracy estimates. Source coverage, implicit-reference recovery and unversioned object identity remain limitations.

Isolation depends on which entities and relations the graph includes (Table 3). Resolution 3 (2023) illustrates the distinction. It explicitly invokes the Environmental Protocol, including Article 7′s prohibition on mineral-resource activities other than scientific research and Article 25′s review provisions, and calls on governments to dispel the misconception that the Treaty or Protocol expires in 2048 (Antarctic Treaty Consultative Meeting, 2023). These references are retained in the separate founding-document layer. The Resolution names no earlier numbered instrument and states no amendment, replacement or revocation of one. Its wording and purpose closely follow Resolution 6 (2016), but this continuity alone does not establish a direct citation or operative relationship (Antarctic Treaty Consultative Meeting, 2016). It is therefore isolated within the numbered-instrument network while remaining explicitly connected to the foundational rules it reaffirms.

**Table 3: Resolution 3 (2023) under three network definitions.** Isolation means zero incoming and outgoing edges within the specified layer, not absence of institutional connections. Founding-document references are recorded separately and do not change the reported isolate counts for the numbered-instrument graphs.

| Network | Classification | Interpretation |
|---|---|---|
| Direct links among numbered instruments | Isolated | No recovered incoming or outgoing link to another numbered instrument. |
| Operative actions among numbered instruments | Isolated | No recovered operative action involving another numbered instrument. |
| References including founding documents | Connected | Explicit references to the Treaty and Protocol, including Protocol Articles 7 and 25. |

Lineages were connected components of the action graph when direction was ignored, and instruments without an action edge were retained as isolates. Chain length counted instruments on the longest directed path, rather than elapsed years. Deterministic rules over instrument subject titles classified site plans, historic-site listings, protected-area system rules, Secretariat and administrative matters, visitor matters, and other subjects. Site keys retained namespaces, with SSSI numbers mapped to ASPA numbers by the title-key implementation. A component in which at least half the instruments had site keys was labelled as the protected-area system. Otherwise, its most frequent normalised title was used when shared by at least two instruments and 30% of the component, with the earliest instrument's title as a fallback. These labels described the component rather than establishing that every connected rule concerned one policy object.

The maintenance share was the proportion of rules that act on at least one other rule dated no later, grouped by adoption in five-year periods with 2020–2025 pooled as the final period and shown only for periods with at least eight rules, on both layers and for all rules as well as without site rules (plans, historic-site listings and protected-area system rules). For rules adopted through 2014 and followed to the end of the record in 2025, an event was a distinct year in which later rules acted on the rule, so that several actions in one calendar year counted once. Repeat action was the share of rules already acted on in $k$ years that were acted on in a further year. Pathway archetypes used the same events, ended at the first year containing a replacement or revocation, and grouped rules into those with no recovered later action by 2025, those replaced at the first event, those acted on once without replacement, and those acted on in two or in three or more years. No later action meant none was recovered before the observation endpoint.

The route machine classified the first operative event after adoption as replacement or revocation, or as revision when a later rule amended the rule, and followed revised rules to a later replacement. Transition shares used visits to the source state as their denominators. These were descriptive sequences, not a fitted Markov process. Topic spread was the Shannon entropy, in natural units, of the six subject-title classes among the rules adopted in a period.

Time to first later action was measured from adoption and summarised with Kaplan–Meier estimates, censoring instruments without a recovered action in 2025. Fixed-window comparisons used adoption cohorts with sufficient follow-up, covering five- and ten-year summaries, a comparison of first actions at lags 0–12 years for instruments adopted in 1995–2012, and ten-year paper-link comparisons for instruments adopted in 1991–2015. We reported Wilson 95% intervals for proportions. These summaries remain conditional on documentary coverage and extraction.

## Routes through the meeting

The CEP–plenary comparison used database agenda assignments for working papers of ATCM 22–47. A paper belonged to the CEP route when all its recorded assignments were CEP items and to the plenary route when all were ATCM items. Papers appearing on both agendas, lacking an exclusive assignment, or not named in the report were excluded. The refreshed submission-stage machine prioritised source-reviewed outcomes at the primary meeting, so a later instrument association did not itself count as adoption at submission. Other outcomes grouped noting, referral, deferral, withdrawal and unclear dispositions.

Rules were assigned to a route when their linked working papers with adoption or later ties all belonged to that route. The comparison followed rules adopted in 1998–2014 through 2025. It summarised their first and subsequent actions and the share of revisions or replacements made by an instrument linked to a new paper of the same route. Sensitivity comparisons restricted route attribution to stronger links, to same-meeting links or to links other than site-title matches, restricted the cohort to rules adopted in 2005–2014, or excluded site rules. Agenda route, instrument form and topic were observed together, so route differences were interpreted as associations in the reconstructed record.

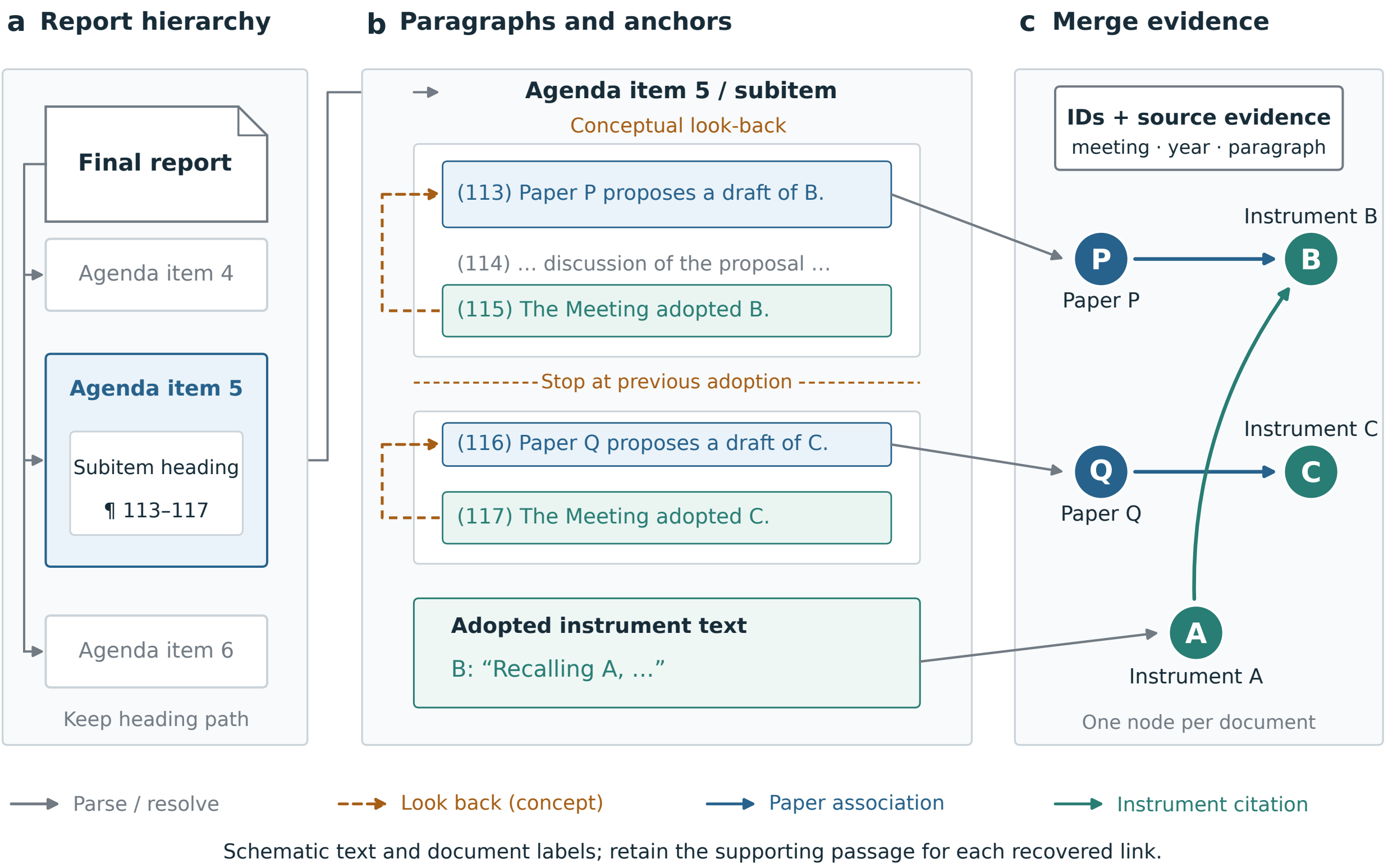

**Figure S2: From report structure to documentary links.** Agenda items, subheadings, and numbered paragraphs define the report context. Schematic passages distinguish proposals, discussion, and adoption anchors. Dashed orange arrows illustrate look-back from an adoption anchor to preceding discussion, with the previous adoption marking a conceptual search boundary. A similar bounded search exists in an archived support-linking pilot, which does not feed the current graph. Grey arrows indicate parsing and identity resolution, and blue arrows show paper–instrument associations P to B and Q to C. The teal arrow A to B represents B's citation of A in its adopted text. Repeated mentions resolve to shared document identities while source evidence is retained. Sentences and lettered documents are illustrative, not observed cases.

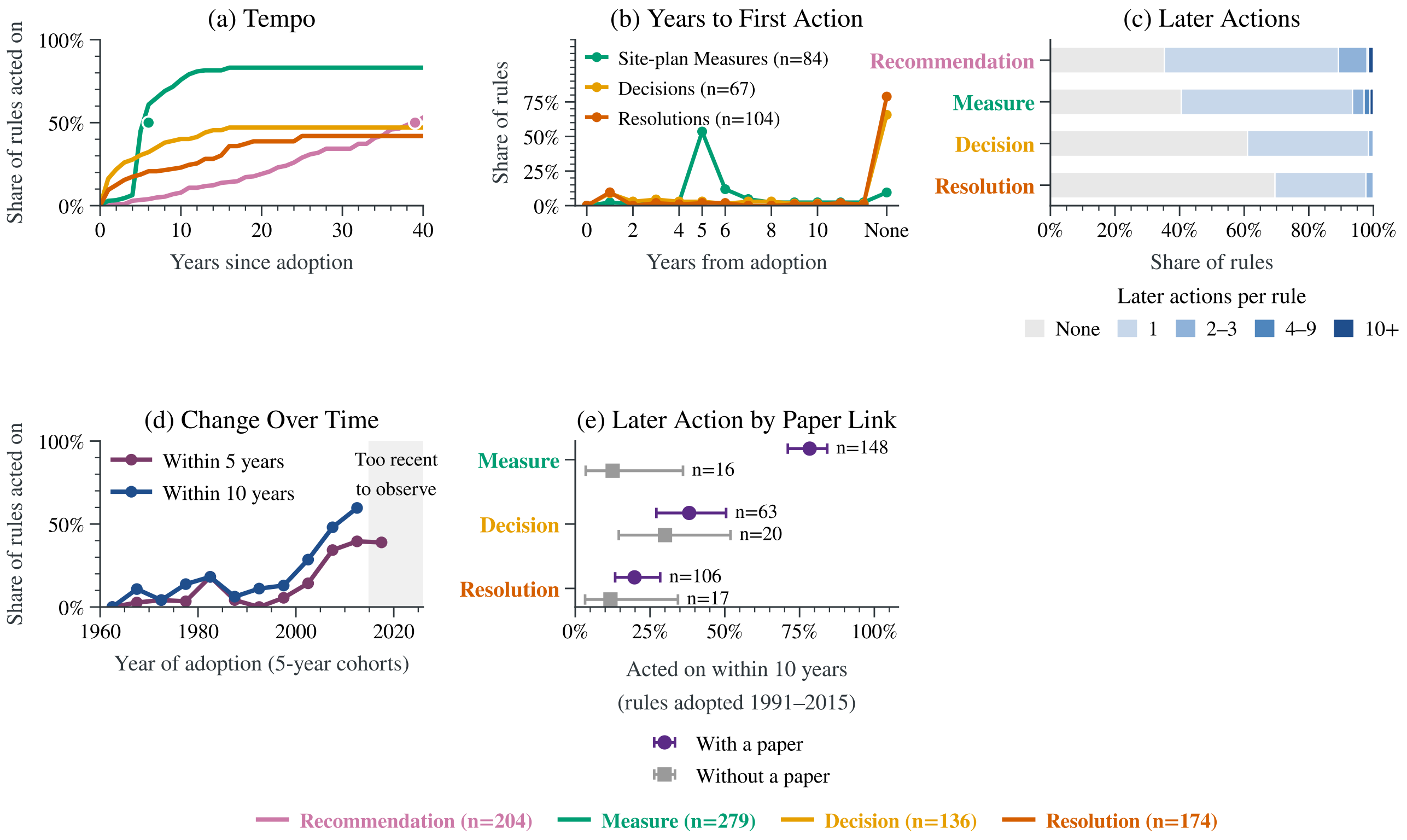

**Figure S3: Further structure of the rule graph (operative actions).** (a) Years from adoption to the first later operative action, by instrument form (Kaplan–Meier, censored in 2025, with dots marking medians). (b) First operative action for site-plan Measures, Decisions and Resolutions adopted in 1995–2012, with first actions displayed at lags 0–12 years. The last point is the share without a recovered action by year 12. (c) Number of later operative actions received, by form. (d) Shares acted on within five and ten years, by adoption cohort, with insufficiently observed cohorts shaded. (e) Operative action within ten years for instruments adopted in 1991–2015, grouped by whether a same-meeting paper association is recovered (Wilson 95% intervals). Instrument actions are operative provisions rebuilt from the instrument readings, and paper flags use the reconciled association table. These comparisons are descriptive associations, not effects of submissions.

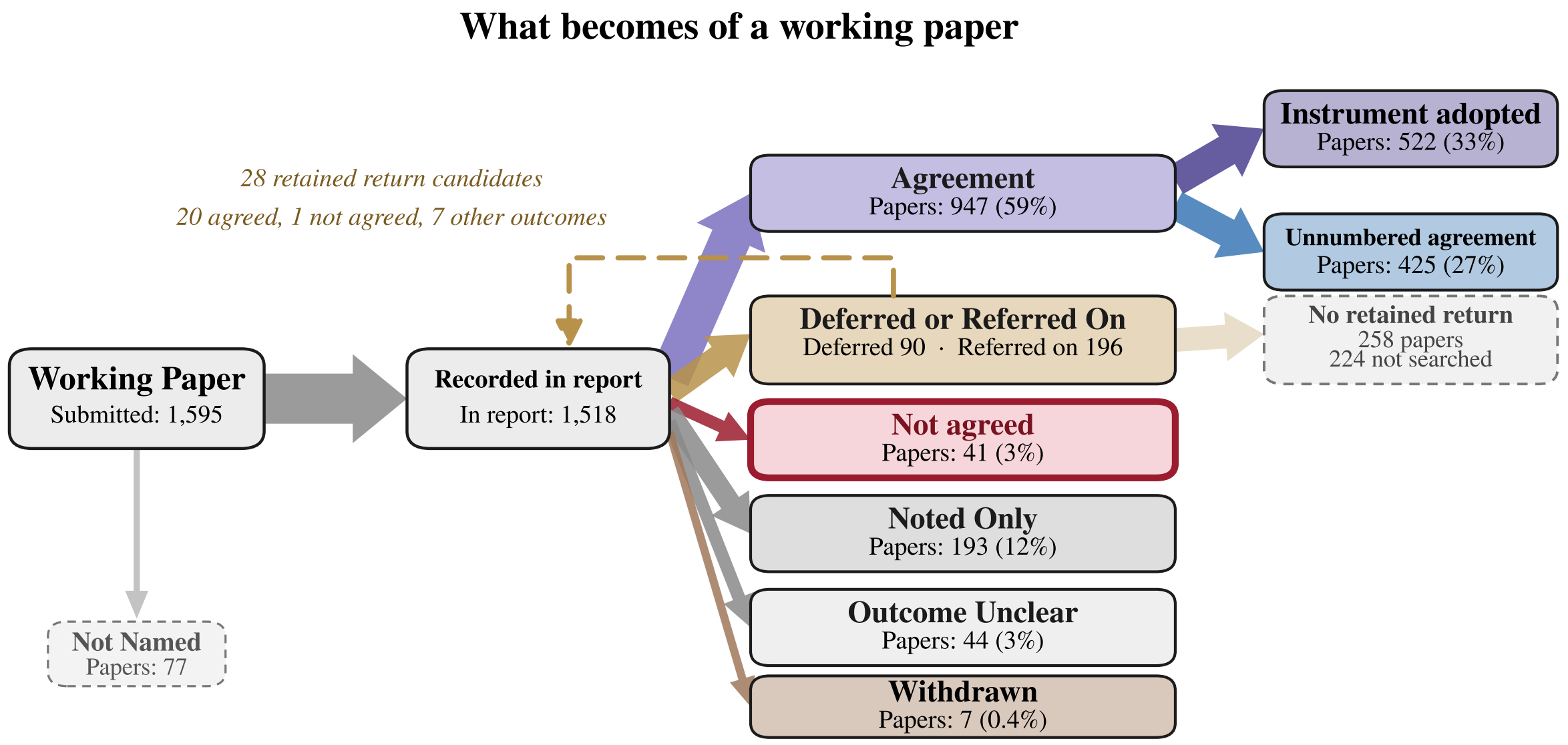

**Figure S4: Working-paper outcomes.** The 1,595 working papers of ATCMs 16–47 include 522 with a recorded adoption of a numbered instrument and 425 endorsements or agreements without a numbered instrument, giving 947 agreement dispositions. Other dispositions are referral or deferral (286), noting (193), collective non-agreement (41), mixed or unclear (44), not identified in the report (77) and withdrawal (seven). Paper–instrument associations alone do not establish adoption or override the disposition recorded at the paper's meeting. The retained successor search covers 62 of the 286 referred or deferred papers and supplies 28 candidate returns in later papers, comprising 27 assessed as the same proposal and one possible match. The successor papers contain 20 agreement dispositions, one non-agreement and seven other outcomes. The remaining 224 referred or deferred papers were not covered by this historical search, so the absence of a retained return does not establish that a proposal never returned. Later operative actions on the rules are shown in Figure S7 f and Figure S5. Outcomes combine model extractions and model-based source review, not independent human validation.

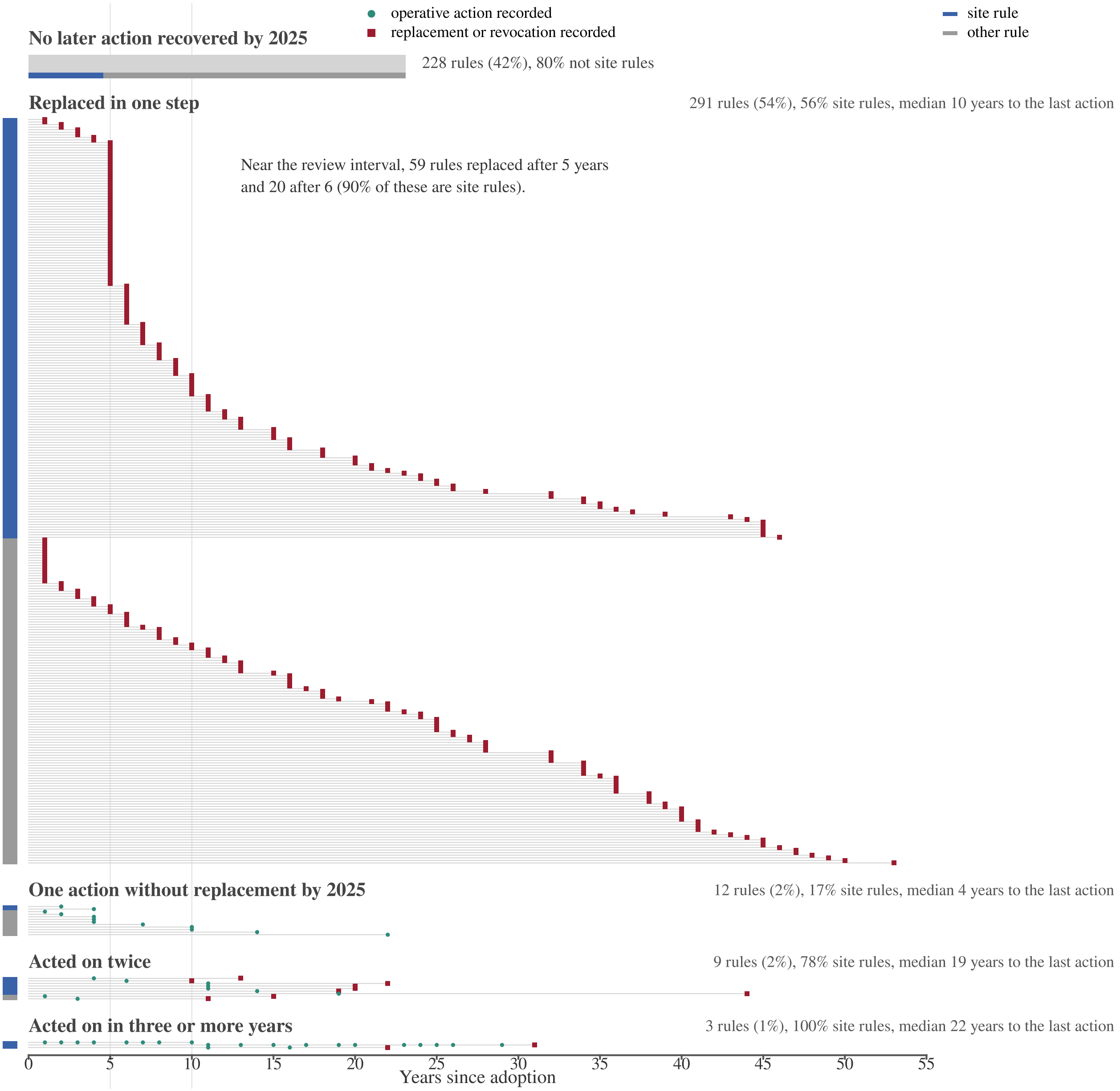

**Figure S5: Pathways of instruments adopted through 2014.** Each line represents one of 543 instruments. Marks identify calendar years with a recovered operative action, and a replacement-or-retirement mark ends the plotted path. Groups contain 228 with no recovered action, 291 replaced or retired at the first event, twelve with one action without replacement, nine with actions in two years and three in three or more years. Multiple actions in one year form one event. The left stripe marks title-classified site instruments. This stopping convention includes component replacement and administrative retirement and does not establish whole-instrument legal lifespans. Follow-up ends in 2025 and differs by adoption year.

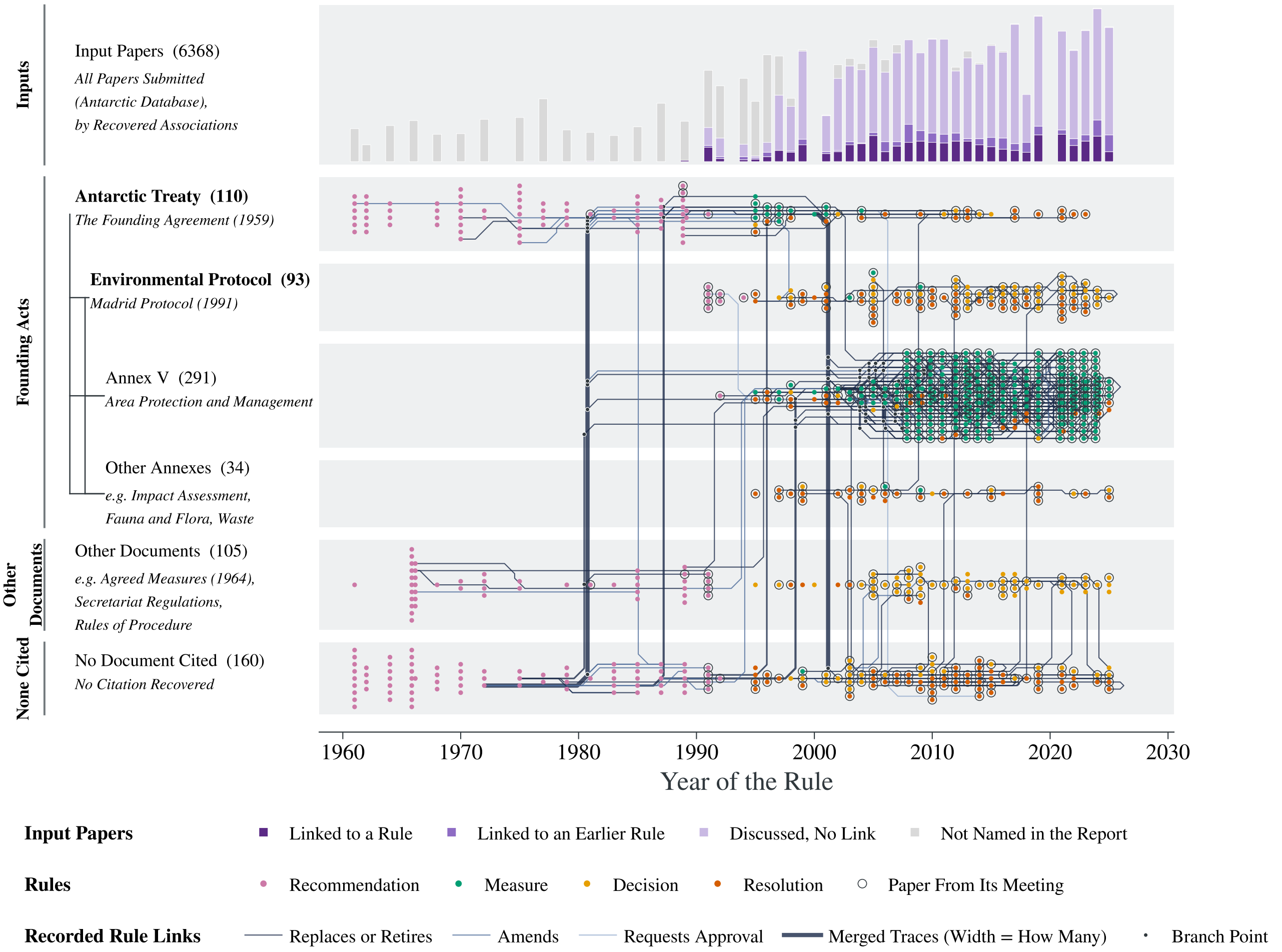

**Figure S6: Instrument histories and paper associations.** Rows indicate the deepest recorded founding-document citation. Dots are instruments placed at adoption, rings identify 536 instruments with a same-meeting paper association, and lines show a selected operative predecessor for each acting instrument. Selection prioritises a predecessor in the same citation row and then the most recent target, so the drawing does not include every operative edge. The top row summarises associations and recorded report mentions in the 6,368-paper inventory. Mention status uses the completed source review for the working-paper cohort and the historical readings for other papers. Citation categories do not establish legal authority, and associations do not establish influence or proposal adoption.

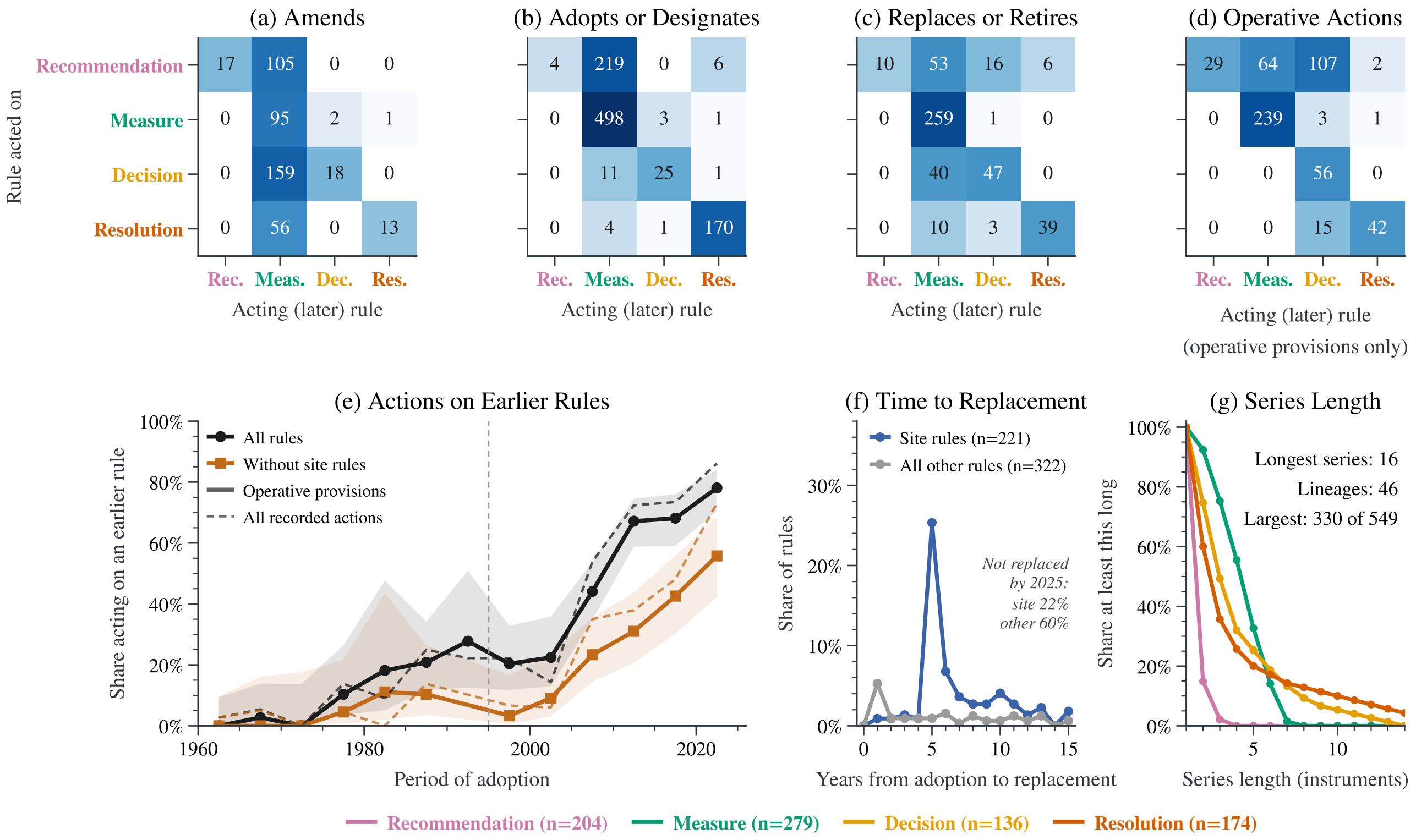

**Figure S7: Who acts on whom, and how maintenance accumulates.** Rules are Recommendations, Measures, Decisions and Resolutions. (a to c) Recorded actions by the kind of the affected rule (rows) and the acting rule (columns), as labelled in the instrument readings. Many are preamble recitals of earlier acts, including all 167 records on Decision 1 (2002), the renaming of protected areas. (d) The operative layer contains 558 actions, comprising 500 replacements or retirements, 56 amendments and two approval requests. Shading uses one log scale for all four panels. (e) The share of rules that act on another rule dated no later, grouped by adoption in five-year periods with 2020–2025 pooled as the final period, for all rules and without plans, listings and site-system rules. Solid lines, with Wilson 95% intervals, count operative actions, and dashed lines include the broader recorded layer as a sensitivity comparison, not an accuracy bound. The vertical line marks 1995. (f) Years from adoption to the first operative replacement or revocation, for site rules and all other rules adopted through 2014 and observed through 2025. The note gives the share without a recovered replacement by 2025. (g) Series length, the number of instruments on the longest chain of operative actions ending at a rule, by kind. Only the recorded graph excludes one backward-dated edge and both directions of one reciprocal same-year pair to make it acyclic, whereas the operative graph has no corresponding deletions.

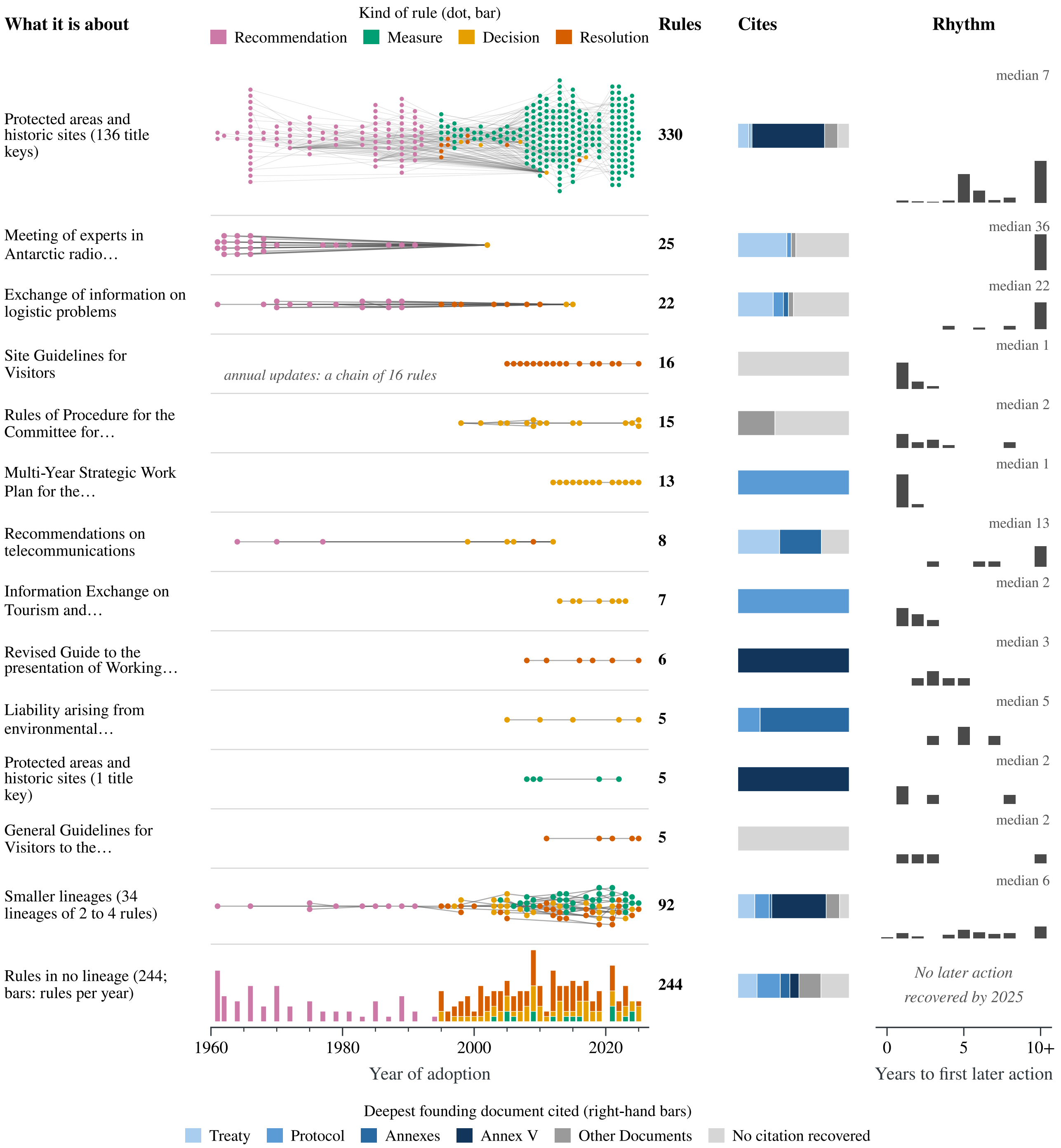

**Figure S8: Subjects and histories of connected instruments.** The 46 lineages contain 549 instruments, alongside 244 isolates. Larger components occupy individual rows, while 34 components of two to four instruments are pooled. Dots mark adoption, colours identify instrument form and lines show operative relations. Labels summarise titles, not single policy objects, and any site counts in labels are title-derived keys, not a validated census of physical sites. Bars show the deepest recorded founding-document citations, and the final panel shows time to first recovered action. The largest mixed component contains 330 instruments and includes bulk status retirements. Visitor site guidelines form the longest recovered chain, with 16 Resolutions, and the strategic work plan has 13 Decisions. Citation categories and action timing do not independently establish legal authority or whole-instrument lifespan.

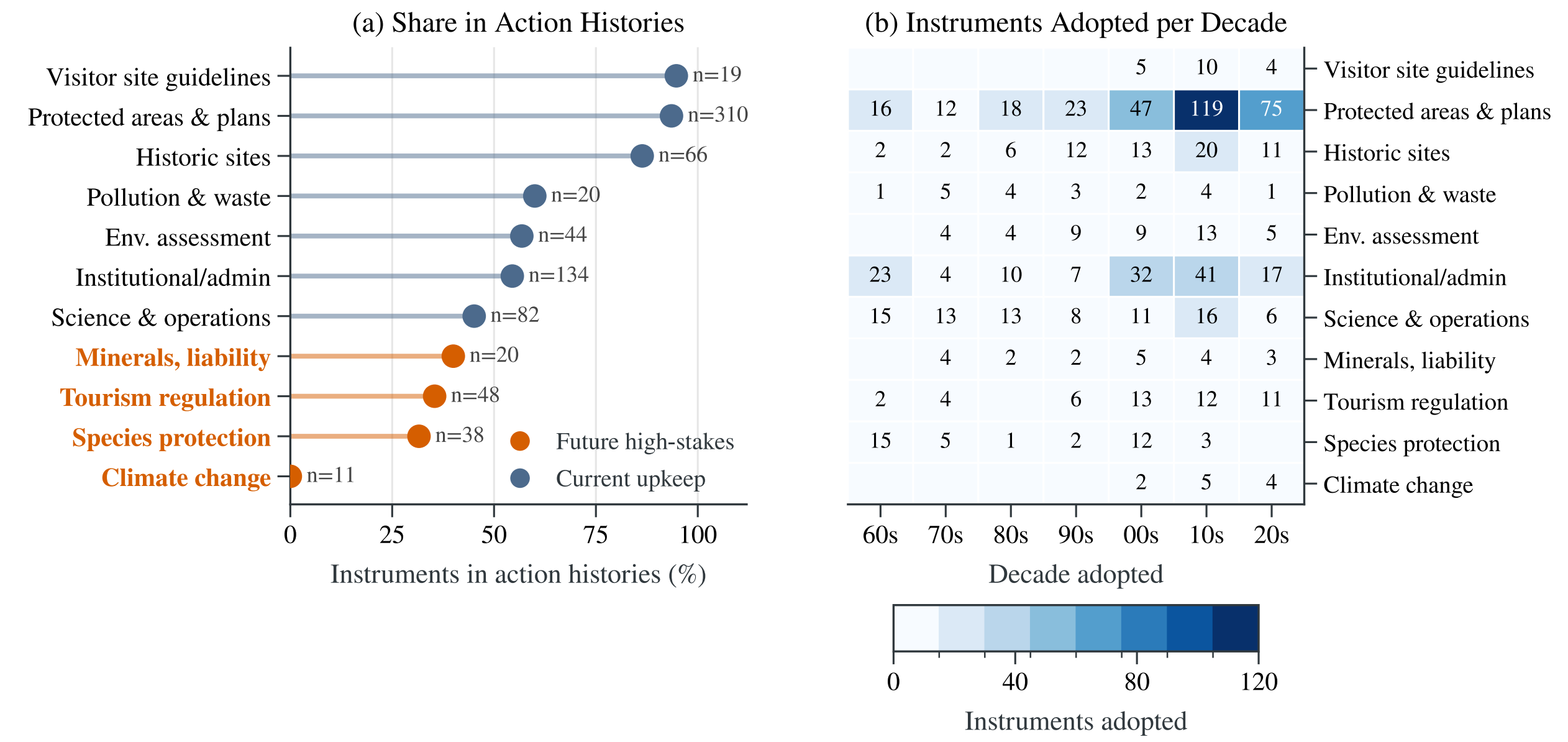

| | 60s | 70s | 80s | 90s | 00s | 10s | 20s |
|---|---|---|---|---|---|---|---|
| Visitor site guidelines | | | | | 5 | 10 | 4 |
| Protected areas & plans | 16 | 12 | 18 | 23 | 47 | 119 | 75 |
| Historic sites | 2 | 2 | 6 | 12 | 13 | 20 | 11 |
| Pollution & waste | 1 | 5 | 4 | 3 | 2 | 4 | 1 |
| Env. assessment | | 4 | 4 | 9 | 9 | 13 | 5 |
| Institutional/admin | 23 | 4 | 10 | 7 | 32 | 41 | 17 |
| Science & operations | 15 | 13 | 13 | 8 | 11 | 16 | 6 |
| Minerals, liability | | 4 | 2 | 2 | 5 | 4 | 3 |
| Tourism regulation | 2 | 4 | | 6 | 13 | 12 | 11 |
| Species protection | 15 | 5 | 1 | 2 | 12 | 3 | |
| Climate change | | | | | 2 | 5 | 4 |

*Components contain at least two instruments linked by operative actions.*
*Domains follow title keywords. Other (n=1) omitted. Adoption counts from the 1960s.*

**Figure S9: Action histories by policy domain (descriptive).** (a) Share of instruments belonging to a component of at least two instruments joined by operative actions, including administrative retirement, with domain sample sizes. (b) Adoption counts per decade. Domains use keyword rules on official titles. Orange identifies domains judged likely to face future tests, a grouping that is interpretive rather than estimated from the graph. Component membership does not establish recurring substantive revision within a domain, since a bulk status review can connect otherwise separate subjects.

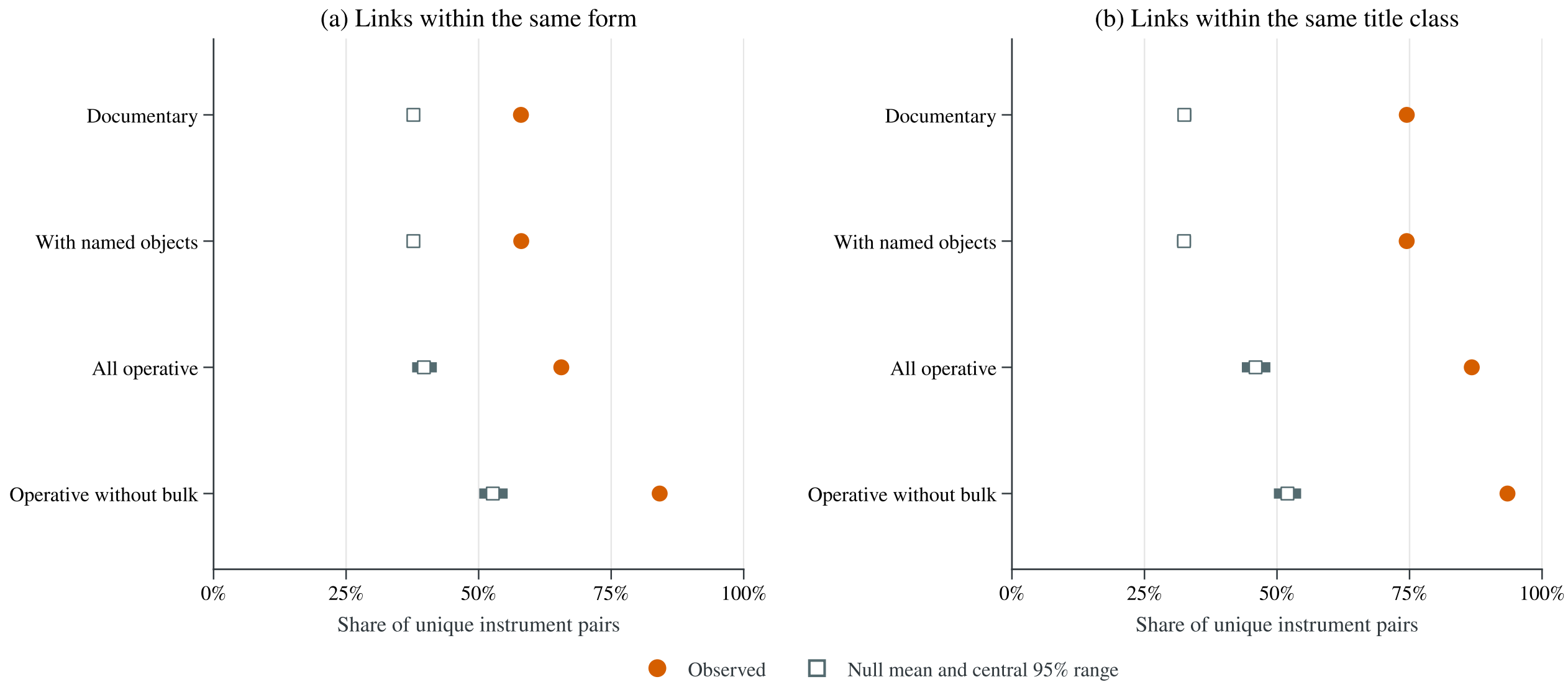

**Figure S10: Specialisation beyond matched network opportunities.** Points show observed within-form and within-title-class shares, and comparison intervals show central 95% ranges over two seeded chains of 500 degree-, chronology- and target-age-matched networks. Documentary, named-object-expanded, full operative and non-bulk operative views retain all 793 instruments. Additional form- or title-conditioned comparisons are reported in the text. Title classes describe instrument functions, not independently coded policy agendas. Null variation does not quantify extraction error, and the comparison does not identify a causal effect of legal form or consensus.